\documentclass[aps,pra,twocolumn,letterpaper,floatfix,superscriptaddress,nofootinbib,10pt,preprintnumbers]{revtex4-1}

\usepackage{amsmath,amssymb}
\usepackage{graphicx,ulem}
\usepackage{enumitem}
\usepackage[pdftex,dvipsnames,usenames]{xcolor}
\usepackage[colorlinks=true,urlcolor=blue,citecolor=blue,linkcolor=blue]{hyperref}

\newcommand{\ket}[1]{\left|#1\right\rangle}
\newcommand{\bra}[1]{\left\langle #1\right|}

\begin{document}
\preprint{PHYSICAL REVIEW A \textbf{97}, 023845 (2018)}

\title{Geometrical picture of photocounting measurements}

\author{O. P. Kovalenko}
	\affiliation{Department of Optics, Palack\'y University, 17. listopadu 12, 771 46 Olomouc, Czech Republic}
	\affiliation{Institute of Physics, National Academy of Sciences of Ukraine, Prospect Nauky 46, 03028 Kiev, Ukraine}
	\affiliation{Physics Department, Taras Shevchenko National University of Kiev, Prospect Glushkova 2, 03022 Kiev, Ukraine}

\author{J. Sperling}
	\affiliation{Clarendon Laboratory, University of Oxford, Parks Road, Oxford OX1 3PU, United Kingdom}

\author{W. Vogel}
	\affiliation{Institut f\"ur Physik, Universit\"at Rostock, Albert-Einstein-Stra\ss{}e 23, 18059 Rostock, Germany}

\author{A. A. Semenov}
	\affiliation{Institut f\"ur Physik, Universit\"at Rostock, Albert-Einstein-Stra\ss{}e 23, 18059 Rostock, Germany}
	\affiliation{Institute of Physics, National Academy of Sciences of Ukraine, Prospect Nauky 46, 03028 Kiev, Ukraine}

\date{\today}

\begin{abstract}
	We revisit the representation of generalized quantum observables by establishing a geometric picture in terms of their positive operator-valued measures (POVMs).
	This leads to a clear geometric interpretation of Born's rule by introducing the concept of contravariant operator-valued measures.
	Our approach is applied to the theory of array detectors, which is a challenging task as the finite dimensionality of the POVM substantially restricts the available information about quantum states.
	Our geometric technique allows for a direct estimation of expectation values of different observables, which are typically not accessible with such detection schemes.
	In addition, we also demonstrate the applicability of our method to quantum-state reconstruction with unbalanced homodyne detection.
\end{abstract}


\maketitle

\section{Introduction}
\label{Sec:Intro}
	An intriguing property of the quantum measurement principle is the non-deterministic nature of this process, often referred to as the collapse of the wave function.
	The standard approach to describing physical measurement outcomes in the quantum domain is given in terms of observables, which are Hermitian operators acting on the Hilbert space of pure quantum states.
	Furthermore, the seminal Born rule states that eigenvalues of these observables represent the possible outcomes of the related measurements with probabilities obtained by averaging the projector for the corresponding eigenstate \cite{Born}.
	Being one of the most fundamental principles of quantum physics, Born's rule was experimentally confirmed, for example, through quantum interferences \cite{Sinha}.
	In addition, it has been considered in connection with other essential concepts of quantum physics, such as entanglement \cite{Harris}.

	The photoelectric detection of light \cite{Mandel, Kelley} is a prominent example of a quantum measurement, which became one of the most important tools to describe quantum-optical experiments.
	In the ideal case, the corresponding measurement outcome is the number of detected photons.
	This formally means that a detector measures the observable $\hat{n}$, the photon-number operator.
	Consequently, an ideal experiment records data which counts $n$ photons with a probability $p_n$.
	However, a realistic description has to model different imperfections properly, requiring one to modify the measured observable.

	Another frequently occurring problem in quantum optics consists in finding the expectation value of an observable $\hat{B}\equiv B(\hat{n})$, which is a function of $\hat{n}$.
	For example, one might be interested in moments or normal-ordered moments of the photon-number operator $\hat{n}$, which are important for the characterization of nonclassical properties of light \cite{Agarwal1992, Kuehn}.
	Another relevant function is the normal-ordered exponent of the photon-number operator, whose expectation value yields the Cahill-Glauber quasi-probability distribution \cite{Cahill1, Cahill2}.
	This feature is used, for example, for the reconstruction of the quantum state of light via unbalanced homodyne detection \cite{Wallentowitz, Mancini, Banaszek1999}.
	These examples demonstrate that the description of expectation values of operator functions is vital for the characterization of quantized radiation fields.
	
	In the case of ideal photodetection, the problem of finding $\langle\hat{B}\rangle$ is straightforwardly resolved by applying the rule
	\begin{align}
		\label{Eq:Mean1}
		\langle\hat{B}\rangle=\sum\limits_{n=0}^{\infty}p_n \bra{n}\hat{B}\ket{n}=\sum\limits_{n=0}^{\infty}p_nB(n),
	\end{align}
	where $\ket{n}$
	is the number state, the eigenstate of $\hat{n}$ to the eigenvalue $n$.
	Therefore, we can directly estimate the value of $\langle\hat{B}\rangle$ from the experimentally obtained photon-number distribution $p_n$.
	Again, the scenario of imperfect detection requires extensive modifications.

\begin{figure}[htb]
\includegraphics[width=1\linewidth]{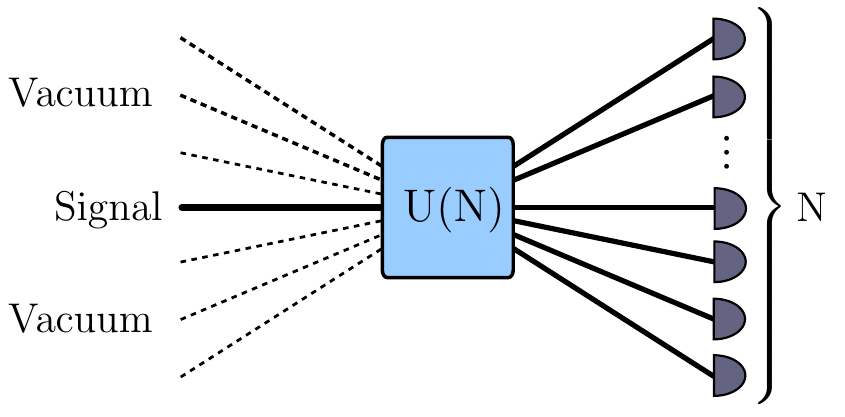}
\caption{\label{Fig:ArrayDetector}
	An array detector consists of a set of $N$ on-off detectors.
	An incident signal is split into multiple beams with equal intensities, represented by a unitary $U(N)$, which are individually measured with on-off detectors.
	Dashed lines indicate the $N-1$ vacuum inputs.
}
\end{figure}

	Realistic detection scenarios and an efficient data analysis necessitate a profound revision of the description of quantum measurements to include imperfect devices.
	For instance, detection losses and dark counts exist in all practical photodetection processes.
	Moreover, discriminating adjacent photon numbers is another challenging task when employing commercially available detectors.
	In this regard, alternative schemes have to be considered, such as click-counting detectors \cite{Silberhorn2007}.
	The corresponding measurement layout is based on the spatial \cite{ArrayDetectors1,ArrayDetectors2,ArrayDetectors3,ArrayDetectors4} or temporal \cite{LoopDetectors1,LoopDetectors2,Rehacek} splitting of the incident light into several optical modes with fewer photons (see Fig. \ref{Fig:ArrayDetector}).
	This renders it possible to detect the resulting modes with devices with a limited photon-number resolution, specifically, on-off detectors which can only react to the presence or absence of photons.
	Here the typical assumption is that if $n$ out of $N$ detectors of the array jointly register clicks, one can expect that this corresponds to $n$ photons in the initial light field.
	However, this naive picture does not apply to many practical cases and as such, the actual click-counting distribution $\varrho_n$ can significantly differ from the photon-number distribution $p_n$ \cite{Sperling2012,Sperling2013,Sperling2015,Sperling2016}.
	One possibility to circumvent this problem and to find a general expectation value $\langle\hat{B}\rangle$ based on Eq. \eqref{Eq:Mean1} may consist in a reconstruction of the true photon-number distribution $p_n$.
	Although this presents an ill-posed (inversion) problem, some methods have been introduced which are very beneficial for particular cases \cite{Rehacek,Perina,Harder2014}.
	Still, a general method to access the desired expectation values is missing so far.

	In this paper, we introduce an alternative method for obtaining the expectation values of functions of the photon number from the experimentally accessible click-counting statistics.
	Our technique is based on the more general finding that Born's rule for general types of measurements yields a clear geometric interpretation.
	This is achieved by applying techniques from analytic geometry \cite{Dubrovin,Misner}.
	As an example of practical relevance, we study the click-counting detection as a measurement procedure for which we can not ascribe a Hermitian operator whose eigenvalues correspond to the measurements outcomes.
	Following Holevo \cite{Holevo}, the corresponding observables in such a case are referred to as generalized observables.
	Our geometrical interpretation of Born's rule enables us to determine an operator decomposition, similar to the positive operator-valued measure (POVM) expansion, for such generalized observables.
		
	The paper is organized as follows.
	In Sec. \ref{Sec:Geometry}, we elaborate the notion of generalized observables and their geometrical structure.
	These results are exemplified through an application to array detectors in Sec. \ref{Sec:AD}.
	In Sec. \ref{Sec:UHD}, the developed theory is further applied to the problem of quantum-state reconstruction.
	A summary and concluding remarks are given in Sec. \ref{Sec:Conclusions}.

\section{Geometry of Born's rule}
\label{Sec:Geometry}

	Let us consider a measurement procedure which is described with the set of operators $\{\hat{\Pi}_n:n\in\mathcal{I}\}$, defining the POVM.
	These operators are positive-semidefinite and satisfy the relation $\sum_{n\in\mathcal{I}}\hat{\Pi}_n{=}\hat{1}$. 
	Since we are interested in the specific class of measurements related to the photoelectric detection of light, we restrict our consideration to such operators $\hat{\Pi}_n$, which are functions of the photon-number operator $\hat{n}$.
	Furthermore, for the sake of simplicity, we will restrict ourselves to countable or finite index sets.

	The probability for the $n$th measurement outcome is described by Born's rule as
	\begin{align}
		\label{Eq:BornRule}
		\varrho_n={\rm Tr}(\hat{\varrho}\,\hat{\Pi}_n),
	\end{align} 
	where $\hat{\varrho}$ is the density operator.
	Let $\hat{B}$ represent a given observable, which is a function of photon-number operator $\hat{n}$.
	Our aim is to formulate a rule which generalizes the relation \eqref{Eq:Mean1} such that the expectation $\langle\hat{B}\rangle$ will be expressed in terms of the probabilities $\varrho_n$.

\subsection{Observables and states}

	The operation of Born's rule in Eq. \eqref{Eq:BornRule} can be considered as a scalar product of $\hat{\varrho}$ and $\hat{\Pi}_n$, which is known as the Hilbert-Schmidt (HS) scalar product and defined as
	\begin{align}
		\langle\hat{\varrho}\, ,\hat{\Pi}_n\rangle_{\rm HS}={\rm Tr}(\hat{\varrho}\,\hat{\Pi}_n).
	\end{align}
	This structure enables one to formally consider observables as elements of a vector space, which is in our case the linear space of HS operators.
	In such geometrical terms, each quantum state can be identified with an element of the dual space, 	that is, a linear functional $\langle\hat{\varrho}\, ,\cdot\rangle_{\rm HS}$ which maps an observable $\hat{B}$ to the number $\langle\hat{B}\rangle=\langle\hat{\varrho}\, ,\hat{B}\rangle_{\rm HS}$, the expectation of this observable.\footnote{The quantum-state functional has to satisfy two conditions:	(i) $ \langle\hat{\varrho}\, ,\hat{g}^\dag\hat{g}\rangle_{\rm HS}\geq 0~\forall \hat{g}$ and (ii) $ \langle\hat{\varrho}\, ,\hat{1}\rangle_{\rm HS}=1$.}
	In addition, the density operator $\hat{\varrho}$ itself is also an element of the HS vector space for the scenario under study.

\begin{figure}[htb]
	\includegraphics[width=0.6\linewidth,clip=]{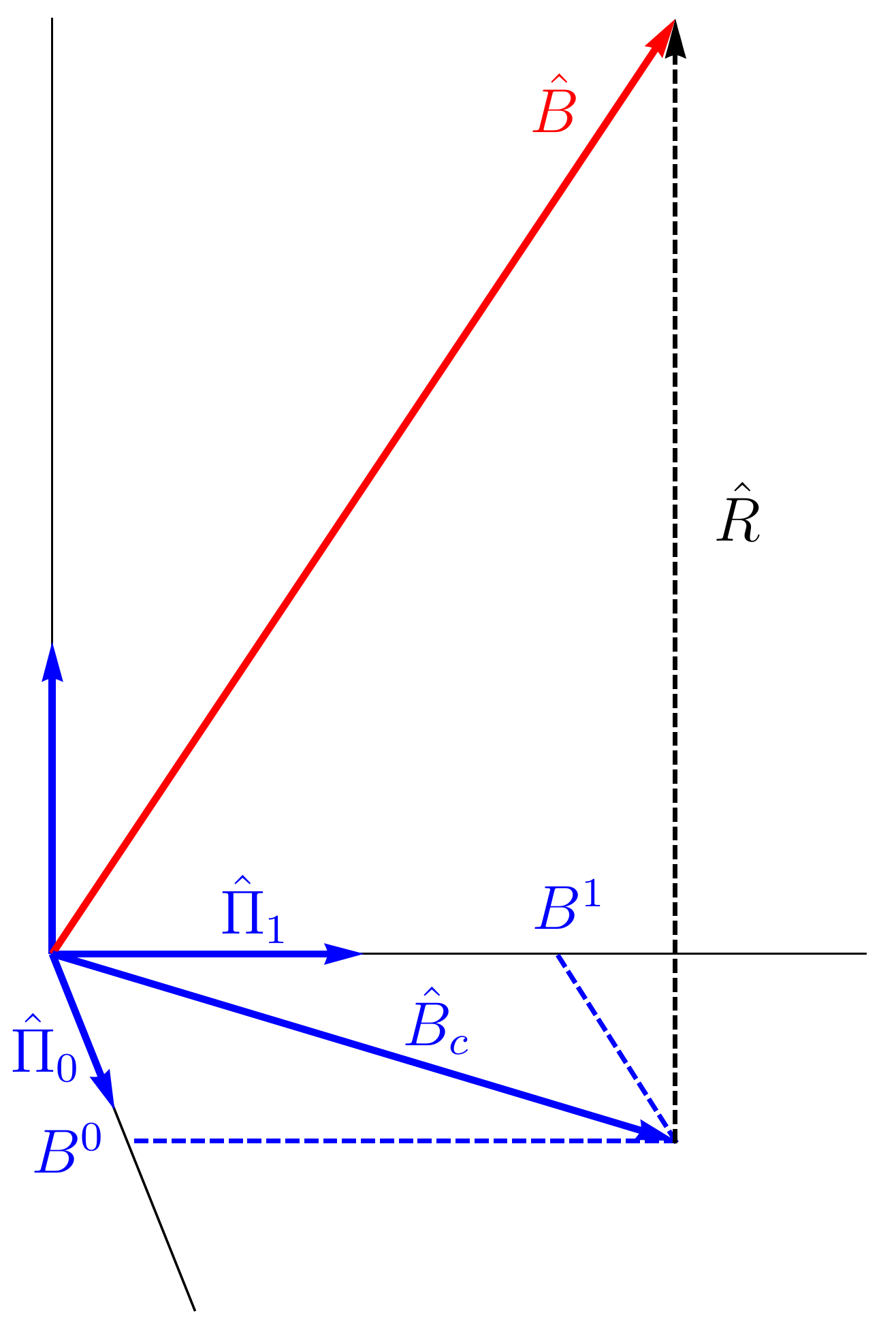}
	\caption{\label{Fig:POVM}
		Schematic presentation of the observable $\hat{B}$ in the POVM basis $\{\hat{\Pi}_0,\hat{\Pi}_1\}$.
		The vector representation of the observable $\hat{B}$ is the sum of the vector $\hat{B}_{c}=\sum_{n=0}^{1} B^n\hat{\Pi}_n$ and the orthogonal completion $\hat{R}$, according to Eq. \eqref{Eq:Expansion}.
	}
\end{figure}

	The POVM  $\{\hat{\Pi}_n:n\in\mathcal{I}\}$ can be considered as a basis for a subspace of observables.
	Note that the POVM, in general, does not span the entire space of HS operators which happens, for example, for the finite index set $\mathcal{I}$ in the case of array detectors as discussed later.
	The observable $\hat{B}$ can be expanded in this basis
	\begin{align}
		\label{Eq:Expansion}
		\hat{B}=\sum\limits_{n\in\mathcal{I}} B^n\hat{\Pi}_n+\hat{R}
	\end{align}
	(see also the schematic presentation in Fig. \ref{Fig:POVM}).
	Here $\hat{R}$ is the orthogonal completion of the operator $\hat{B}$ to the POVM basis to account for the contribution of the operator which is not spanned by the POVM under study.
	This yields for such an orthogonal completion, 
	\begin{align}
		\label{Eq:OrthogonalCompletion}
		\langle\hat{R}\,,\hat{\Pi}_n\rangle_{\rm HS}={\rm Tr}(\hat{R}\,\hat{\Pi}_n)=0
	\end{align}
	for all $n\in\mathcal I$.
	The symbols $B^n$ denotes the contravariant coordinates of the observable $\hat{B}$.
	Note that in analytic geometry, superscripts indicate a specific element and not a power \cite{Dubrovin, Misner}, which is adopted in this paper.
	
	Averaging over the relation \eqref{Eq:Expansion}, i.e., applying the quantum-state functional $\langle\hat{\varrho}\, ,\cdot\rangle_{\rm HS}$, we find the rule for the estimation of the expectation value $\langle\hat{B}\rangle$,
	\begin{align}
		\label{Eq:ExpectEstim}
		\langle\hat{B}\rangle=\sum\limits_{n\in\mathcal{I}} B^n\varrho_n+\langle\hat{R}\rangle.
	\end{align}
	Here $\langle\hat{R}\rangle$ can be understood as a systematic error due to the orthogonal complement to the given POVM.
	For practical applications of this relation, we also need to describe how to calculate the contravariant coordinates $B^n$ of the observable $\hat{B}$.

	First, we consider the special case of an observable $\hat C$ for which the values $C^n$ can be considered as the outcomes of the given measurement procedure.
	For example, $C^n=n$ can represent the number of registered clicks from an array detector.
	In this case, Eqs. \eqref{Eq:Expansion} and \eqref{Eq:ExpectEstim} are given by
	\begin{align}
		\hat{C}=\sum\limits_{n\in\mathcal{I}} C^n\hat{\Pi}_n,~		
		\langle\hat{C}\rangle=\sum\limits_{n\in\mathcal{I}} C^n\varrho_n,
	\end{align}
	respectively.
	This clearly implies that $\langle\hat{C}\rangle$ is the expectation of the generalized observable related to the considered measurement.
	Thus $\hat{C}$ can be considered as the operator representation of the measured observable.
	However, its eigenvalues coincide with the measurement outcomes solely in the case when the POVM is orthonormal, i.e. $\langle\hat\Pi_n,\hat\Pi_m\rangle_{\rm HS}=\delta_{mn}$ for all $m,n\in\mathcal I$, where $\delta$ denotes the Kronecker symbol.
	A different rule has to be developed for obtaining more general contravariant coordinates $B^n$.
	Let us emphasize that for the general case, we have to make a distinction between the contravariant coordinate and the measurement outcome.

\subsection{Contravariant operator-valued measure}

	In order to find such a rule, we introduce a method based on the geometrical description.
	For this purpose, we consider which information about the quantum state can be extracted from the given measurement.
	We can interpret Born's rule \eqref{Eq:BornRule} as a geometrical relation $\varrho_n=\langle\hat{\varrho}\,,\hat{\Pi}_n\rangle_{\rm HS}$, where the probabilities $\varrho_n$ can be interpreted as the covariant coordinates of the density operator $\hat{\varrho}$.
	This means we can use the fact that there exists a dual operator basis $\{\hat{\Pi}^n:n\in\mathcal{I}\}$, termed the \textit{contravariant operator-valued measure} (COVM), such that
	\begin{align}
		\hat{\varrho}=\sum\limits_{n\in\mathcal{I}}\varrho_n\hat{\Pi}^n+\hat{O},
	\end{align}
	where $\hat{O}$ is the orthogonal completion to the density operator $\hat{\varrho}$.
	To ensure that the relation \eqref{Eq:ExpectEstim} holds true, the POVM and the COVM operators, being dual bases to each other, satisfy the orthogonality relation,
	\begin{align}
		\label{Eq:Orthogonality}
		\langle\hat{\Pi}^n,\hat{\Pi}_m\rangle_{\rm HS}={\rm Tr}(\hat{\Pi}^n\hat{\Pi}_m)=\delta^n_m.
	\end{align}

	Moreover, any basis is completely characterized by the covariant metric tensor \cite{Dubrovin, Misner}
	\begin{align}
		g_{nm}=\langle\hat{\Pi}_n,\hat{\Pi}_m\rangle_{\rm HS}={\rm Tr}(\hat{\Pi}_n\hat{\Pi}_m).
		\label{MetricTensorCov}
	\end{align}
	In the case of an orthonormal basis, we have $g_{nm}=\delta_{nm}$, which is no longer true in the general case.
	To construct the dual basis for the general scenario, the contravariant metric tensor $g^{nm}$, the inverse to $g_{nm}$, can be computed as
	\begin{align}
		\sum\limits_{k\in\mathcal{I}}g^{nk}g_{km}=\delta^n_m.
		\label{MetricTensorContr}
	\end{align}
	Therefore, the COVM elements can be straightforwardly obtained via
	\begin{align}
		\hat{\Pi}^n=\sum\limits_{m\in\mathcal{I}} g^{nm}\hat{\Pi}_m.
		\label{COVM}
	\end{align}
	By such a construction, the POVM and the COVM necessarily satisfy the orthogonality relation \eqref{Eq:Orthogonality}.

	This COVM technique allows us to operate with (generalized) observables in the same manner as vectors are handled in analytic geometry.
	In particular, the contravariant coordinates of the observable $\hat{B}$ are given by
	\begin{align}
		\label{Eq:ContravariantCoord}
		B^n=\langle\hat{\Pi}^n,\hat{B}\rangle_{\rm HS}={\rm Tr}(\hat{\Pi}^n\hat{B}).
	\end{align} 
	This approach generalizes and includes the previously considered case of the observable $\hat{C}$, for which contravariant coordinates $C^n$ are identical to the outcomes of the measurement procedure.
	Beyond the following examples, we also apply this technique to photocounting detection including losses \cite{Mandel,Kelley} (where $\hat{R}=0$) in Appendix \ref{App:PD}.

	In addition, we can now also define the covariant coordinates of observables,
	\begin{align}
		\label{Eq:CovCoord}
		B_n=\langle\hat{B},\hat{\Pi}_n\rangle_{\rm HS}={\rm Tr}(\hat{B}\hat{\Pi}_n).
	\end{align} 
	Further, applying the formula  \eqref{Eq:ExpectEstim}, the contravariant coordinate reads
	\begin{align}
		\label{Eq:RisingIndeces}
		B^n=\sum\limits_{m\in\mathcal{I}}g^{mn}B_m,
	\end{align}
	which represents a raising of indices, frequently used in analytic geometry.

\subsection{Discussion}

	As previously mentioned, a geometrical structure naturally appears in quantum theory when describing Born's rule using the HS product.
	In this interpretation, each observable can be considered as a vector in a given POVM basis.
	Quantum states are presented by elements of the corresponding dual space.
	Furthermore, let us mention that in general, the COVM (the dual basis) is not a positive semidefinite one.
	Thus, the COVM does not represent a physical measurement.

\begin{figure}[htb]
	\includegraphics[width=0.75\linewidth]{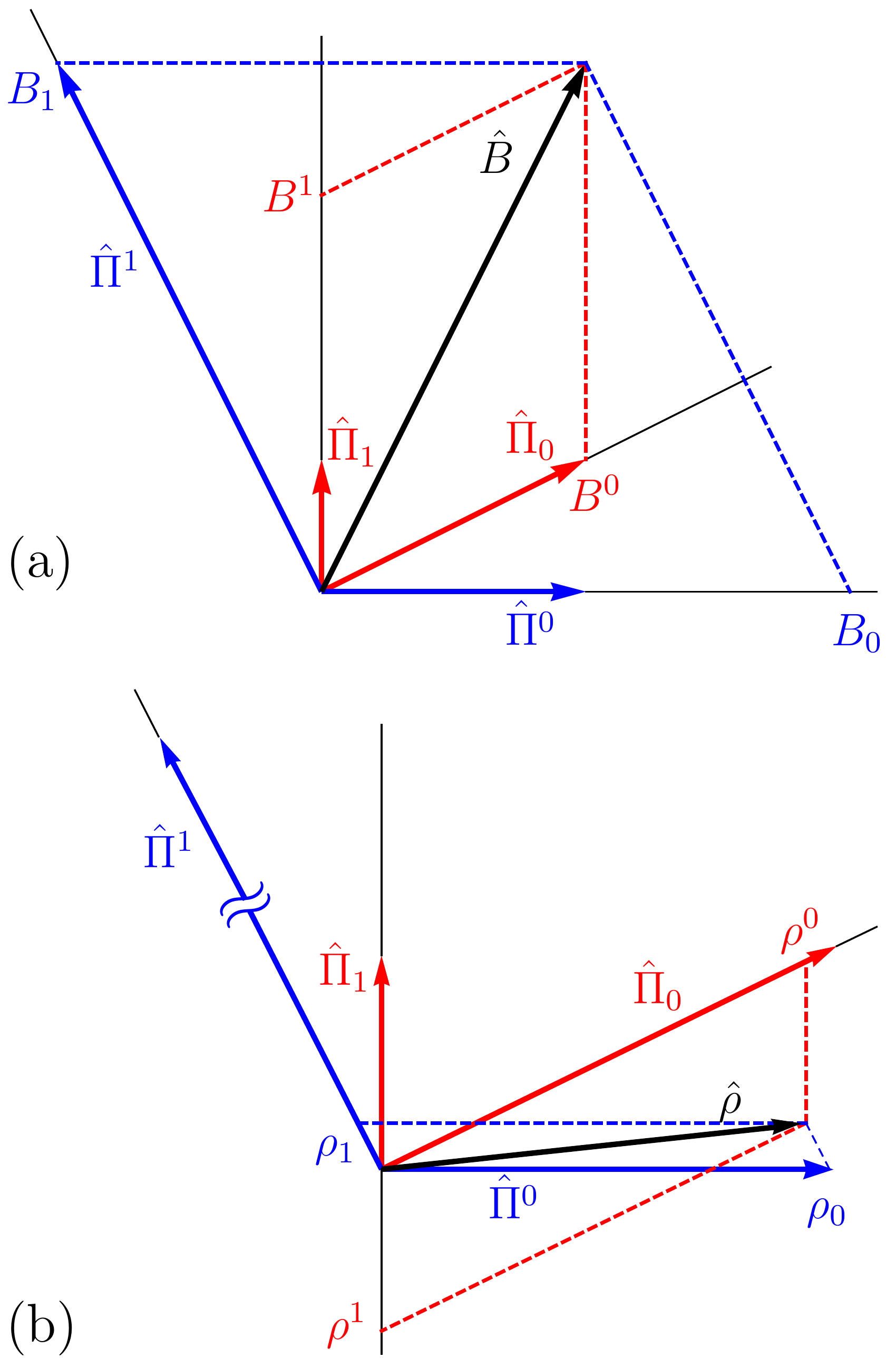}
	\caption{\label{Fig:COVM}
		Graphical presentation of POVM and COVM bases, Eqs. \eqref{Eq:POVM_Example} and \eqref{Eq:COVM_Example}, respectively, for $\eta=0.5$.
	 (a) The observable $\hat{B}=\ket{0}\bra{0}+2\ket{1}\bra{1}$ and (b) the density operator $\hat{\varrho}=0.9\ket{0}\bra{0}+0.1\ket{1}\bra{1}$  with the corresponding covariant and contravariant coordinates are shown.
		The coordinate $\rho^1$ has a clear negative value.
	}
\end{figure}

	In Fig. \ref{Fig:COVM}(a), we show an example of a POVM
	\begin{align}\label{Eq:POVM_Example}
		\hat\Pi_0=|0\rangle\langle 0|+(1-\eta)|1\rangle\langle 1|
		\quad ,~\quad
		\hat\Pi_1=\eta|1\rangle\langle 1|
	\end{align}
	together with its COVM,
	\begin{align}\label{Eq:COVM_Example}
		\hat\Pi^0=|0\rangle\langle 0|
		\quad ,~\quad
		\hat\Pi^1=-\frac{1-\eta}{\eta}|0\rangle\langle 0|+\frac{1}{\eta}|1\rangle\langle1|.
	\end{align}	
	For instance, $|0\rangle$ and $|1\rangle$ could be the ground and excited state of a two-level system, respectively. 
		The excitation is detected with a quantum efficiency $\eta\in[0,1]$.
	It is easy to verify that the orthogonality ${\rm Tr}(\hat\Pi^m\hat\Pi_n)=\langle\hat\Pi^m,\hat\Pi_n\rangle_{\rm HS}=\delta_{n}^m$ is satisfied for all $m,n\in\mathcal I=\{0,1\}$.
	The coordinate axes shown in Fig. \ref{Fig:COVM} point into the perpendicular directions $|0\rangle\langle 0|$ and $|1\rangle\langle 1|$.
	While the POVM is positive, the negative component of its COVM can be clearly seen.

	The geometric interpretation of Born's rule for a generalized observable $\hat C$ is summarized by the following two statements:
		(i) The measurement outcome is given by the contravariant coordinates $C^n$ of the measured observable $\hat{C}$
		and (ii) the probability of the $n$th outcome is given by the corresponding covariant coordinate $\varrho_n$ of the density operator $\hat{\varrho}$.
	Another observable $\hat{B}$ can be expanded in the POVM basis of the observable $\hat{C}$, which results in the rule \eqref{Eq:ExpectEstim} for an approximate estimation of its expectation value.
	In this context, let us mention that a special rule for defining functions of generalized observables can be formulated too (see Appendix \ref{App:Functions}).

	Furthermore, the length of the vector is an important geometrical property.
	In our case, it is given by the HS norm of the corresponding operator
	\begin{align}
		\label{Eq:HS_Norm}
		\|\hat{B}\|_\mathrm{HS}=\sqrt{\langle\hat{B},\hat{B}\rangle_{\rm HS}}=\sqrt{{\rm Tr}(\hat{B}^{[2]})}.
	\end{align}
	Here and in the following, superscripts of the form $[n]$ denote the $n$th power to distinguish them from a component of a coordinate.
	In the scenario under study, the considered operators belong to the HS class, i.e. $\|\hat{B}\|_\mathrm{HS}<\infty$.
	Not all observables satisfy this condition, such as the photon-number operator $\hat{n}$.
	In Sec. \ref{Sec:AD}, we additionally discuss a way to overcome this problem for some physically relevant examples.

	We can consider covariant coordinates of observables [see Eq. \eqref{Eq:CovCoord}] and contravariant coordinates of the density operator $\varrho^n=\langle\hat{\Pi}^n,\hat{\varrho}\rangle_{\rm HS}={\rm Tr}(\hat{\Pi}^n\hat{\varrho})$.
	This allows us to expand the observable and the density operators with these coordinates as
	\begin{align}
		\hat{C}=\sum\limits_{n\in\mathcal{I}}C_n\hat{\Pi}^n,\quad~ \quad\hat{\varrho}=\sum\limits_{n\in\mathcal{I}}\varrho^n\hat{\Pi}_n+\hat{O}.
	\end{align}
	In contrast to $C^n$ and $\varrho_n$, neither $C_n$ nor $\varrho^n$ represents values appearing in the geometrical formulation of Born's rule.
	This means that they do not have a physical meaning in terms of a measured and reconstructed quantity.
	For instance, $\varrho^n$ can take even negative values [see Fig. \ref{Fig:COVM}(b) for an example].
	Yet these new coordinates provide helpful tools for the geometric interpretation of the quantum-physical measurement process.

	In addition, there exist well-known examples of detection schemes beyond the photocounting measurement, such as eight-port \cite{Walker} and heterodyne detection \cite{Helstrom}.
	In this scenario, the POVM can be presented by an uncountable set of projectors on coherent states $\hat\Pi_n=\pi^{-1}|\alpha\rangle\langle\alpha|$  with $n=\alpha\in\mathbb C=\mathcal I$.
	Then $\varrho^n$ is identical to the Glauber-Sudarshan $P$ function \cite{Glauber,Sudarshan} (a not necessarily non-negative phase-space distribution) $\varrho^n=P(\alpha)$ and $\varrho_n$ is the Husimi-Kano $Q$ function \cite{Husimi,Kano} (a necessarily non-negative phase-space distribution) $\varrho_n=Q(\alpha)$.
	We additionally have a vanishing orthogonal completion to the density operator\ $\hat{O}=0$.
	Such a geometrical representation for quantum phase-space distributions has been considered also in Ref. \cite{Berezin}.
	These measurements, which allow for reconstructing the full density operator, are called informationally complete (see Refs. \cite{Busch,DAriano,Prugovecki,Schroeck,Renes}).

\subsection{Singular metrics}
\label{Sec:Singularity}

	The geometrical structure of finite-dimensional subspaces, i.e. $\mathcal I=\{0,\ldots, N\}$ yields a singularity in the covariant metric tensor.
	In order to demonstrate this, one can consider the resolution of unity for the POVM $\sum_{n=0}^{N}\hat{\Pi}_n=\hat{1}$, where $\hat{1}$ is the identity.
	From this property, we can conclude that at least one of the operators $\hat{\Pi}_n$ does not belong to the HS class, which can be seen when expressing one of them, e.g. $\hat{\Pi}_N$ in terms of the other ones $\hat{\Pi}_N=\hat{1}-\sum_{n=0}^{N-1}\hat{\Pi}_n$.
	Evidently, even if $g_{nm}={\rm Tr}(\hat{\Pi}_n\hat{\Pi}_m)$ exists for $n,m=0\ldots N-1$, it cannot exist for $\hat{\Pi}_N$ because $\hat{1}$ is not a HS operator $\|\hat 1\|_\mathrm{HS}=(\sum_{n=0}^\infty 1)^{1/2}=\infty$.
	This means that at least one component of the covariant metric tensor (here $g_{NN}$) yields a singular value.

	To ensure that our method applies to such a scenario as well, let us consider the following.
	To compensate for the singularity of the covariant metric tensor, the contravariant metric tensor $g^{nm}$ satisfies
	\begin{align}
		\label{Eq:ContrMetricN1}
		g^{Nm}=g^{nN}=g^{NN}=0
		\quad\text{and}\quad
		\sum_{k=0}^{N-1}g^{nk}g_{kn}\,=\delta_n^m.
	\end{align}
	This also implies that $\hat{\Pi}^{N}=0$.
	All other components of the COVM $\hat{\Pi}^n$ are linear combinations of the HS-class POVM components $\hat{\Pi}_n$ for $n=0,\ldots, N-1$.
	Therefore, the COVM is represented by HS-class operators only.

\subsection{Systematic error}
\label{Sec:FiniteBasis}

	For a wide class of HS operators $\hat{B}$, $\|\hat{B}\|_\mathrm{HS}<\infty$ [Eq. \eqref{Eq:HS_Norm}], one can quantify the contribution of the orthogonal completion $\hat{R}$.
	In fact, it follows from Eqs. \eqref{Eq:Expansion}, \eqref{Eq:OrthogonalCompletion}, and \eqref{Eq:ContrMetricN1} that the HS norm of the orthogonal completion reads
	\begin{align}
		\|\hat{R}\|_\mathrm{HS}=\sqrt{\|\hat{B}\|_\mathrm{HS}^{[2]}-\sum_{n=0}^{N-1}B_nB^n}.
		\label{Eq:HS_R}
	\end{align}
	This norm yields an upper bound to the systematic error of the evaluation of $\langle\hat{B}\rangle$,
	\begin{equation}
		\langle\hat{R}\rangle^{[2]}\leq
		\langle\hat{R}^{[2]}\rangle\leq
		\|\hat{R}\|_\mathrm{HS}^{[2]}.
		\label{ErrorEst1}
	\end{equation}
	Later, we also demonstrate that the actual errors are usually much smaller than estimated by this worst-case scenario.
	Therefore $\|\hat{R}\|_\mathrm{HS}$ can be considered as a state-independent HS mismatch for the estimation of the expectation value $\langle\hat{B}\rangle$ based on the reconstructed probabilities $\varrho_n$ for an observable $\hat{B}$ based on the measurement of the generalized observable $\hat{C}$ (see the scheme in Fig. \ref{Fig:POVM}).

\section{Application: Click Detectors}
\label{Sec:AD}

	In this section, we consider a typical example of the finite-basis measurement in the infinite-dimensional photon-number space of practical relevance.
	One feature of such measurements is that a clear discrimination of photon numbers, which corresponds to the POVM considered in Appendix \ref{App:PD}, is hardly accessible with the presently available technologies.
	There exist several ways to resolve this problem.
	For example, one way to circumvent imperfect photon-number resolution is to split an initial beam into a number of beams and to detect each of them with an on-off detector, which results in an array of click detectors (see, e.g., Ref. \cite{ArrayDetectors1,ArrayDetectors2,ArrayDetectors3,ArrayDetectors4} and Fig. \ref{Fig:ArrayDetector}).
	A related approach consists in using fiber loop configurations, resulting in arrays of time-bins to be detected \cite{LoopDetectors1,LoopDetectors2,Rehacek}.

	The photocounting equation and the corresponding POVM for array detection schemes have been presented and analyzed in Refs. \cite{Sperling2012, Sperling2013}.
	For $N$ on-off detectors, the POVM is given by the elements
	\begin{align}
		\hat{\Pi}_n ={:}\binom{N}{n}
			\left[\hat1{-}\exp\left(\!{-}\frac{g(\hat{n})}{N}\right)\right]^{[n]} \exp\left(\!{-}\frac{g(\hat{n})}{N}\right)^{[N{-}n]}{:}
		\label{Eq:POVMarray}
	\end{align}
	for $n\in\mathcal I=\{0,\ldots,N\}$ denoting the number of on-off detectors which record a coincidence click.
	Here $g(\hat{n})$ is the detector response function, which can be estimated experimentally \cite{Bohmann}.
	In our case, we assume a linear form $g(\hat{n})=\eta\hat{n}+\nu$, where $\eta$ is the detection efficiency and $\nu$ is the intensity of dark counts \cite{Semenov2008,Lee2005,Pratt1969,Karp1970}.
	Further $:\cdots:$ denotes normal ordering.
	We recall that the superscript $[n]$ indicates the power.
	It is also worth mentioning that in many practical situations, the click statistics Eq. \eqref{Eq:POVMarray} strongly differs from the photocounting statistics Eq. \eqref{POVM_Mandel} (see also Ref. \cite{Sperling2012} for a detailed discussion).

\subsection{Metric tensor and coordinates}
	
	For $\eta=1$ and $\nu=0$, the covariant metric tensor for the POVM elements in Eq. \eqref{Eq:POVMarray} reads
	\begin{equation}
		g_{nm} = \binom{N}{n} \binom{N}{m} \sum_{k=0}^{m}\sum_{l=0}^{n} \binom{m}{k} \binom{n}{l}  \frac{( -1)^{[k+l]}}{1-\frac{(n-l) ( m-k)}{N^{[2]}} }.
		\label{Eq:CovMT}
	\end{equation}
	See Appendix \ref{App:ArrayCov} for the general case $\eta\neq 1$ or $\nu\neq0$.
	As it has been discussed in Sec. \ref{Sec:Singularity}, the covariant tensor of the finite-basis POVM has at least one divergent component, here $g_{NN}=\infty$.
	Consequently, the $(N+1)\times(N+1)$ components of the contravariant metric tensor have to satisfy the conditions in Eq. \eqref{Eq:ContrMetricN1}.
	This implies that we can consider only an ($N\times N$)-dimensional tensor $g^{nm}$, which is inverse to the tensor $g_{nm}$ for $n,m=0,\ldots, N-1$.
	Even though the analytical expression for the specific contravariant metric tensor is unknown, it can be directly obtained numerically.

	Now we aim at reconstructing expectation values of an observable $\hat{B}$, which is a function of the photon-number operator $\hat{n}$, based on the click-counting statistics $\varrho_n$ as stated by the rule \eqref{Eq:ExpectEstim}.
	Therefore, it is enough to determine the covariant coordinates $B_n$  [Eq. \eqref{Eq:CovCoord}], and then calculate the contravariant coordinates via the method of rising indices \eqref{Eq:RisingIndeces}, where $g^{nm}$ is the numerically computed inverse of $g_{nm}$.
	Consequently, the covariant coordinate reads 
	\begin{align}
		B_n=\binom{N}{n}\sum_{k=0}^{n}\binom{n}{k}(-1)^{[n-k]}F_{N;k},
		\label{Eq:CovCoordExpr}
	\end{align}
	where
	\begin{equation}\label{Eq:F}
		F_{N;k}={\rm Tr}\left[\hat{B}:\exp\left(-\frac{(N-k)\hat{n}}{N}\right):\right].
	\end{equation}
	Recall that the index set in all sums is restricted by $\mathcal I=\{0,\ldots,N-1\}$ (see Sec. \ref{Sec:Singularity}).
	In the following, let us apply this method for some important examples of the operator $\hat{B}$ to gather some insight into the photon properties accessible without a full photon-number resolution.

\subsection{Reconstruction of the photon-number statistics}
\label{Sec:PNStat}

	As the first application of our technique, we consider the operator $\hat{B}=\ket{m}\bra{m}$, the projector on a Fock state.
	The expectation of this operator gives the probability of having $m$ photons if the detection was perfect.
	Therefore, the estimation of such operators for different $m$ results in the reconstruction of the photon-number statistics from the click statistics.
	The covariant coordinate of this operator can be explicitly calculated,
	\begin{align}
		B_n=\Big[\ket{m}\bra{m}\Big]_n=
		\binom{N}{n}\frac{n!}{N^{[m]}}
		\left\{\begin{array}{c}
			m\\n
		\end{array}
		\right\},
		\label{TMatrix}
	\end{align}
	where $\left\{\begin{smallmatrix} m\\n\end{smallmatrix}\right\}$ are the Stirling numbers of the second kind.
	For details of the calculation, see also Refs. \cite{Sperling2012,Miatto2016}.

\begin{figure}[htb]
	\includegraphics[width=1\linewidth]{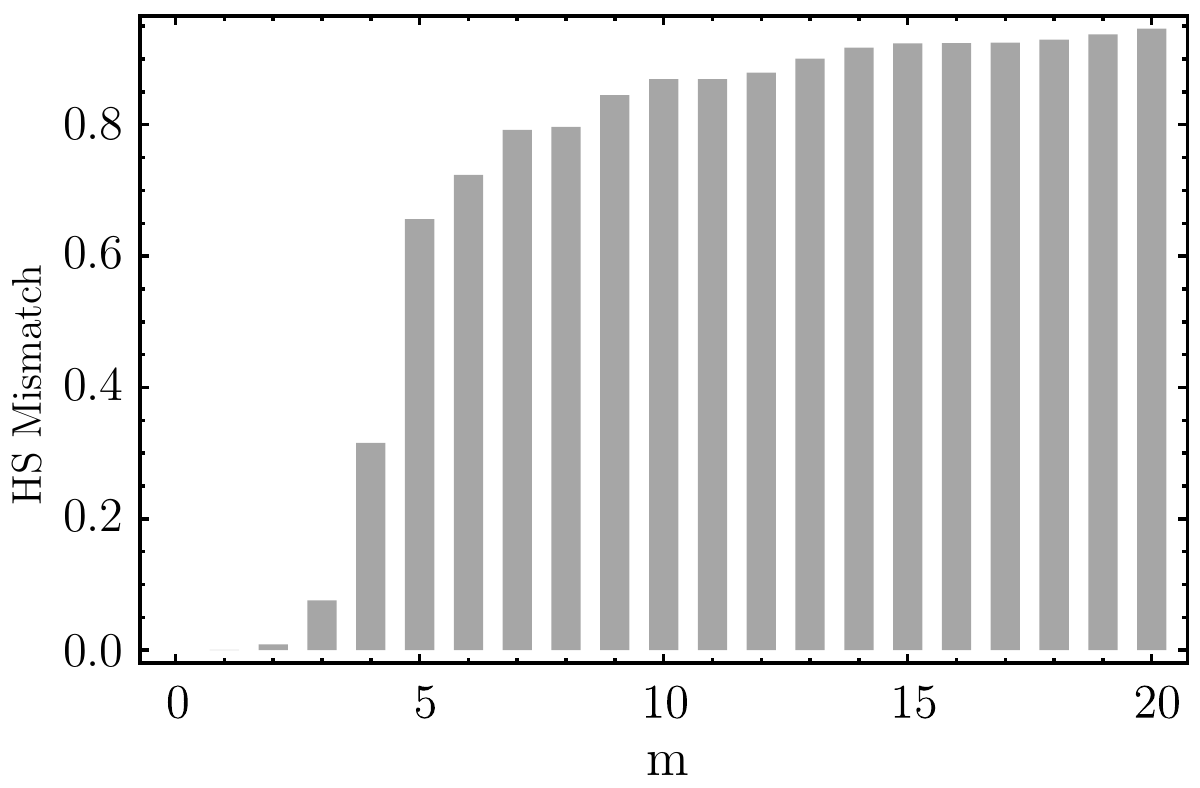}
	\caption{\label{Fig:Fock}
		The HS mismatch of the projectors $\ket{m}\bra{m}$ for different numbers $m$ in the case of $N=10$.
	}
\end{figure}

	The HS mismatch [see Eq. \eqref{Eq:HS_R}] for different projectors $\ket{m}\bra{m}$ and for $N=10$ is shown in Fig. \ref{Fig:Fock}.
	One can see that for $m=0,1,2$, it yields acceptable values.
	For larger $m$, the projectors $\ket{m}\bra{m}$ mostly belong to the orthogonal complement of the basis $\hat{\Pi}_n$.
	This means that the reconstruction of the photon-number statistics from the click statistics without applying additional regularization techniques can completely fail (see Appendix \ref{App:Pseudoinv}).
	Consequently, our method certifies on a quantitative basis, via exact reconstruction errors, that a proper detection theory is necessary to employ array detectors and inversions, as frequently used, have to be handled with great care.

\subsection{Click-counting operator}

	Another fundamental example is the click-counting operator (see Refs. \cite{Sperling2013,Miatto2016}), which reads
	\begin{align}
		\hat{C}=\sum\limits_{n=0}^{N} n\,\hat{\Pi}_n= N\left[1 -: \exp\left(-\frac{\hat{n}}{N}\right): 
		\right].
		\label{Eq:ClickNumberOp}
	\end{align}
	This operators represents the generalized observable corresponding to the click-counting detection.
	Again, the eigenvalues of this operator are different from measurement outcomes.
	However, the contravariant coordinates of this operator are clearly given by
	\begin{align}
		C^n=\langle\hat{\Pi}^n,\hat{C}\rangle_{\rm HS}=n,
	\end{align}
	which completely agrees with the geometrical formulation of Born's rules.

\subsection{Exponents of the photon-number operator}

	Let us consider an exponential function of the photon-number operator $\hat{n}$,
	\begin{align}
		\hat{B}=\exp(-t\hat{n}),
		\label{Eq:ExpOp}
	\end{align}
	where $t\geq 0$ is a fixed parameter.
	Here the $F_{N;k}$ in the expression \eqref{Eq:CovCoordExpr} for covariant coordinates are given by
	\begin{align}
		F_{N;k}=\frac{N}{N-k\exp(-t)}.
		\label{Eq:F_for_Exp}
	\end{align}
	The HS norm of this operator
	\begin{align}
		\|\hat{B}\|_\mathrm{HS}^{[2]}=\frac{1}{1-\exp(-2t)}
	\end{align}
	shows that the operator belongs to the HS class for any positive $t$.

\begin{figure}[htb]
\includegraphics[width=1\linewidth]{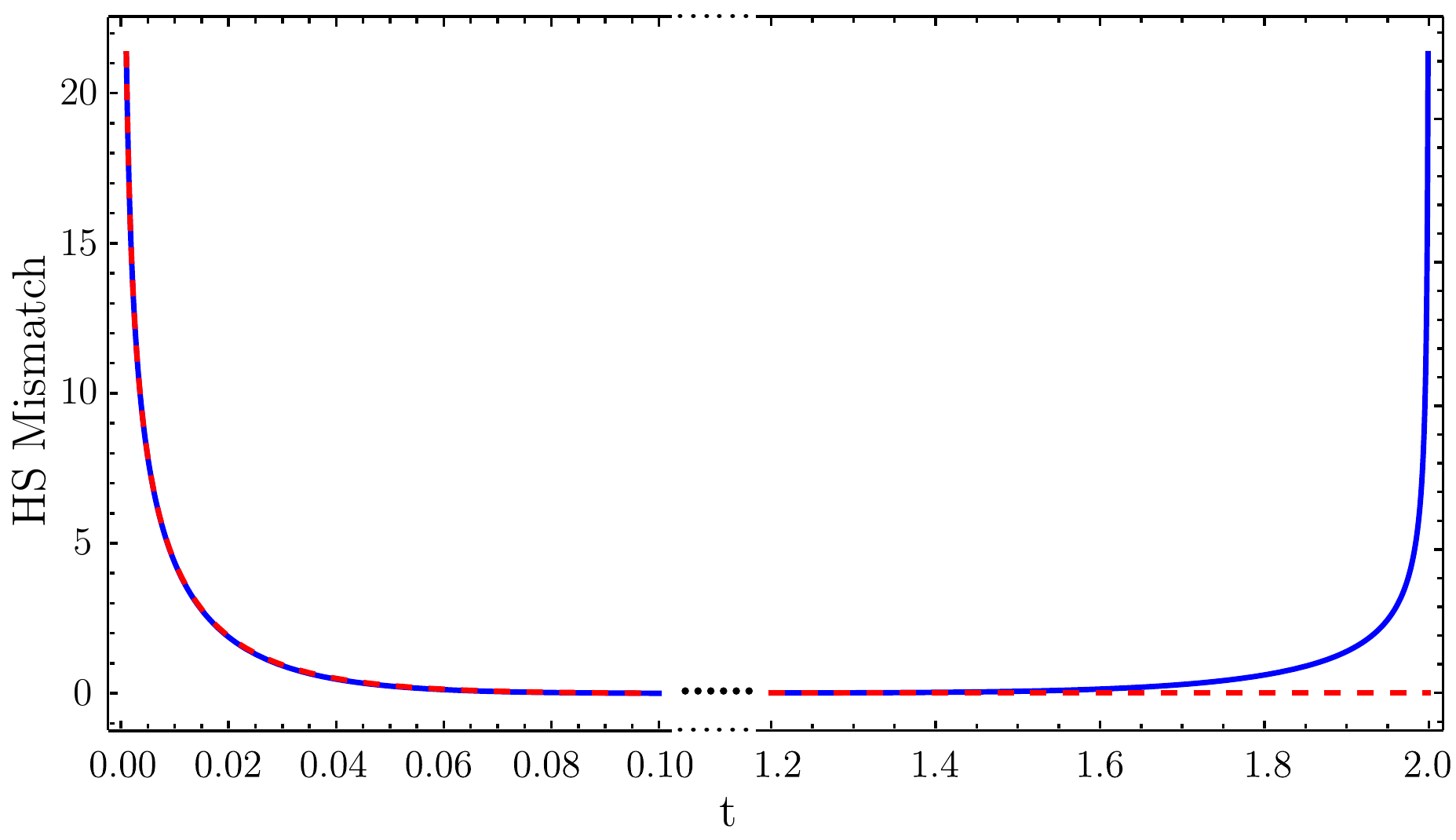}
		\caption{\label{Fig:Exp}
		The HS mismatch $\|\hat{R}\|_\textrm{HS}$ of the operators $\hat{B}=\exp(-t\hat{n})$ (dashed line) and $\hat{B}=:\exp(-t\hat{n}):$ (solid line) vs the parameter $t$ in the case of $N=10$ on-off detectors.
	}
\end{figure}

	Similarly, let us consider the normally ordered operator
	\begin{align}
		\hat{B}=:\exp(-t\hat{n}):,
		\label{Eq:ExpOpN}
	\end{align}
	where $t\in(0,2)$.
	Such an operator plays a crucial role in the quantum-state reconstruction with unbalanced homodyne detection \cite{Wallentowitz} (see also Sec. \ref{Sec:UHD}).
	The numbers $F_{N;k}$ in this case take the form
	\begin{align}
		F_{N;k}=\frac{N}{N-k(1-t)},
		\label{Eq:F_for_ExpN}
	\end{align}
	which yields the HS norm as
	\begin{align}
		\|\hat{B}\|_\mathrm{HS}^{[2]}=\frac{1}{t(2-t)}.
		\label{Eq:HSN_ExpN}
	\end{align}
	This operator belongs to the HS class for $0<t<2$.

	Plots of the HS mismatches $\|\hat{R}\|_\textrm{HS}$ [see Eq. \eqref{Eq:HS_R}] as functions of the parameter $t$ in the case of $N=10$ are presented in Fig. \ref{Fig:Exp} to assess the quality of the reconstruction.
	For very small values of the parameter $t$, both cases of HS mismatches are large, which means that the corresponding operators are mainly spanned by the orthogonal complement of the POVM $\hat{\Pi}_n$.
	With increasing $t$, the mismatches tend to zero.
	In this case, the upper bound of the systematic error for the reconstruction of the expectation values [see Eq. \eqref{ErrorEst1}], is comparably small.
	This behavior holds true for arbitrary large $t$ in the case of $\hat{B}=\exp(-t\hat{n})$.
	However, for $\hat{B}=:\exp(-t\hat{n}):$, the HS mismatch increases when $t$ is close to the value $2$.

\subsection{Moments of the photon-number operator}

	Moments of the photon-number operator $\langle\hat{n}^{[m]}\rangle$ and the normal-ordered moments $\langle:\hat{n}^{[m]}:\rangle$ play a crucial role for the verification of nonclassical properties of light \cite{Agarwal1992, Kuehn}.
	Both operators $\hat{n}^{[m]}$ and $:\hat{n}^{[m]}:$ do not belong to the HS class.
	Thus, the HS mismatch in both cases is undefined.
	Nevertheless, the covariant coordinates in both cases can be obtained by expanding Eqs. \eqref{Eq:F_for_Exp} and \eqref{Eq:F_for_ExpN} in series with respect to $t$.
	The numbers $F_{N;k}$ for the moments are given by
	\begin{align}
		F_{N;k}=\left(x\frac{{\rm d}}{{\rm d} x}\right)^{[m]}(1-x)^{[-1]}\Big\rvert_{x=k/N}
		\label{Eq:Mom}
	\end{align}
	for the case of $\hat{B}=\hat{n}^{[m]}$ and
	\begin{align}
		F_{N;k}=\frac{m!Nk^{[m]}}{(N-k)^{[m+1]}}
		\label{Eq:MomN}
	\end{align}
	for the case of $\hat{B}=:\hat{n}^{[m]}:$.
	These expressions are clearly divergent for $k=N$.
	This means that the covariant coordinates $B_{N}$ are also divergent, which is consistent with the divergence of the metric tensor ($g_{NN}=\infty$).

	Furthermore, such a behavior of the moments has a clear physical explanation.
	Namely, it is impossible to give a state-independent estimation for the moment of photon numbers.
	For an arbitrary state, the probability of appearance of arbitrarily large numbers of photons can be arbitrarily large too.
	It is evidently impossible to estimate the state-independent error without additional information about the state.

	Let us suppose that our state is approximately restricted by a photon number of $M$.
	In this case, we can redefine the photon-number moment $\hat{n}^{[m]}$ as
	\begin{align}
		\hat{B}_{(M)}=\sum_{n=0}^{M}n^{[m]}\ket{n}\bra{n}
		\label{Eq:truncatedMom}
	\end{align}
	and, correspondingly, the normal-ordered photon-number moment $:\hat{n}^{[m]}:$  as
	\begin{align}
		\hat{B}_{(M)}=\sum_{n=m}^{M}\frac{n!}{(n-m)!}\ket{n}\bra{n}.
		\label{Eq:truncatedMomN}
	\end{align}
	This truncation to $M$ photons is indicated by $(M)$ in the index.
	Such truncated versions of moments belong to the class of HS operators.
	In the limit $M\rightarrow\infty$, they approach the operators $\hat{n}^{[m]}$ and $:\hat{n}^{[m]}:$, respectively.

\begin{figure}[htb]
	\includegraphics[width=1\linewidth]{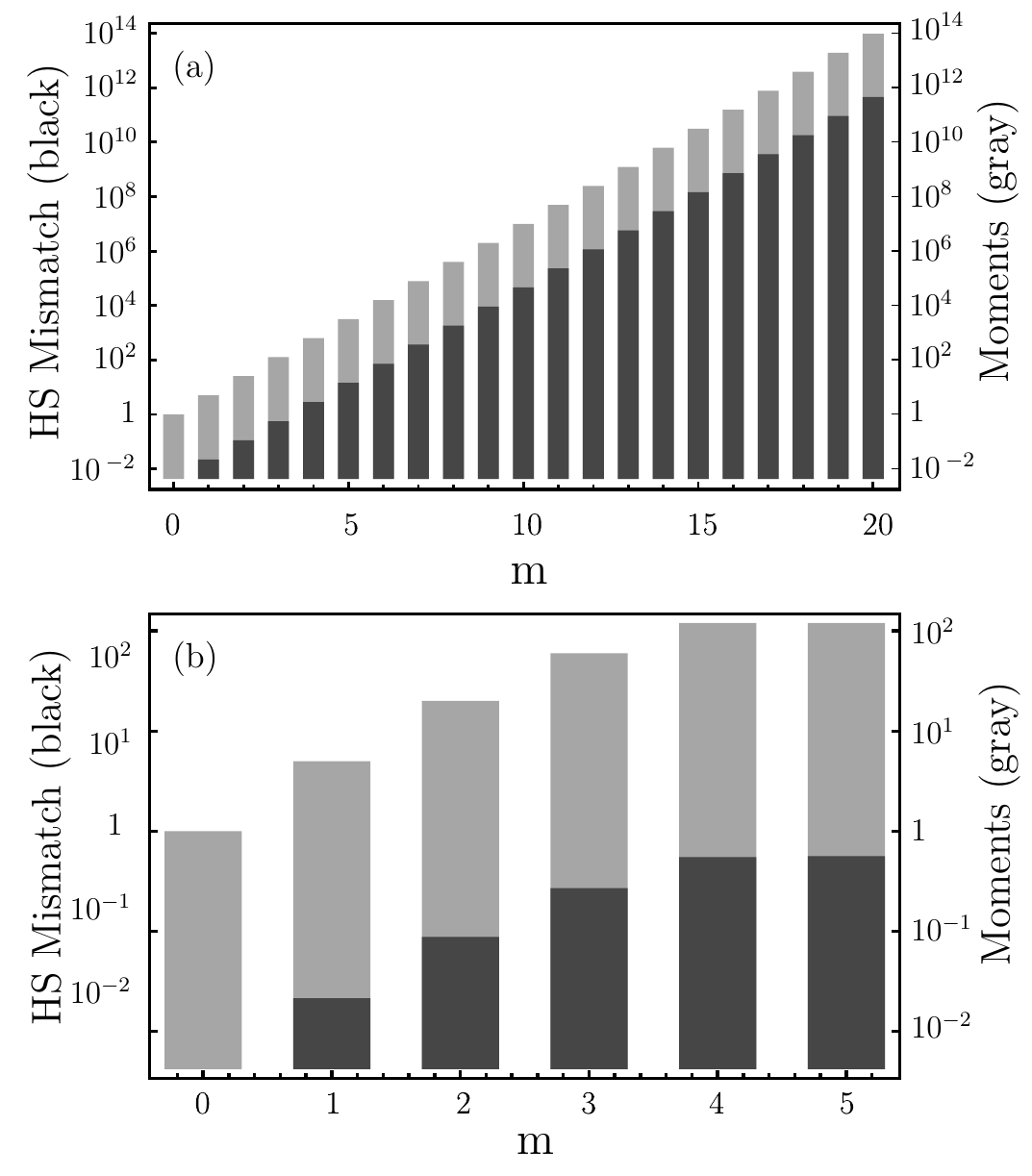}
		\caption{\label{Fig:Mom}
		The HS mismatches $\|\hat{R}\|_\mathrm{HS}$ (black bars) and moments (gray bars) in logarithmic scale for a photon-number state $\ket{5}$ in the case of $N=20$ as a function of the order of moments $m$.
	(a) the truncated moments $\langle\hat{n}^{[m]}\rangle$ [see Eq. \eqref{Eq:truncatedMom}],
		and (b) the truncated normal-ordered moments $\langle:\hat{n}^{[m]}:\rangle$[see. Eq. \eqref{Eq:truncatedMomN}].
	}
\end{figure}

	In Fig. \ref{Fig:Mom}, we present the dependence of the HS mismatch on the number $m$ for the truncated moments and the truncated normal-ordered moments.
	In both cases, the HS mismatches increase not faster than a logarithmic function.
	A similar behavior occurs for the moments themselves.
	In Fig. \ref{Fig:Mom}, we compare the HS mismatch with the value of moments for a photon-number state.
	It can be clearly seen that the relative HS mismatch is about a few orders smaller than the orders of moments.
	This example shows that the reconstruction of moments for a low-intensity light is feasible.

\subsection{Fine estimation of the reconstruction error}

	As it has been discussed in Sec. \ref{Sec:FiniteBasis}, the HS mismatch $\|\hat{R}\|^{[2]}_\mathrm{HS}$ represents the upper bound for the estimation error of an observable $\hat{B}$ based on the click-counting statistics $\varrho_n$.
	As the HS mismatch is a state-independent characteristic of such an observable, its value can be large compared to the expectation value of the observable itself.
	Here we describe a technique which enables one to significantly decrease the estimated error based on additional information about the quantum state under study.

	The idea of fine estimation of errors generalizes the truncation approach, as already demonstrated for the estimation of moments $\langle\hat{n}^{[m]}\rangle$ and normal-ordered moments $\langle:\hat{n}^{[m]}:\rangle$.
	Here we apply this technique to arbitrary operators, which are functions of the photon-number operator $\hat{n}$.
	Specifically, let us suppose that we have to estimate an expectation value of a given observable $\hat{B}$, which is based on the measured click statistics $\varrho_n$.
	This observable can be expanded in terms of projectors on the Fock number states as
	\begin{align}
		\hat{B}=\sum\limits_{n=0}^{\infty}\ket{n}\bra{n}\hat{B}\ket{n}\bra{n}.
		\label{Eq:OpBExp}
	\end{align}
	Presupposing the additional knowledge about the quantum state that, to a good approximation, this states does not have more than $M$ photons, we find that the expectation value of the operator \eqref{Eq:OpBExp} is the same as for the truncated operator
	\begin{align}
		\hat{B}_{(M)}=\sum\limits_{n=0}^{M}\ket{n}\bra{n}\hat{B}\ket{n}\bra{n}.
		\label{Eq:OpBExpMod}
	\end{align}
	Regardless of whether the operator $\hat{B}$ belongs to the HS class, the new operator $\hat{B}_{(M)}$ is a HS operator.

	Such a procedure allows us to estimate the HS mismatch \eqref{Eq:HS_R} for the operator \eqref{Eq:OpBExpMod}.
	Yet it requires additional knowledge about the quantum state in terms of the truncation number $M$.
	This number has to be chosen such that the overall probability to have more than $M$ photons is negligibly small.
	In general, the mismatch of the truncated operator $\hat{B}_{(M)}$, estimating the reconstruction error for the observable, is smaller than the mismatch of the operator $\hat{B}$.
	
	In this section, we applied our geometric approach to measurements which are incomplete and therefore exhibit singularities in their metric tensor.
	Using our techniques, we were able to approximate vital features of the photon-number statistics based on measurements with frequently employed array detectors.
	Moreover, this included a rigorous treatment of reconstruction errors which enable a quantitative assessment of the resulting expectation values.
	Our derived methods are of major relevance for experiments which rely on such reconstruction approaches and have to include appropriate error estimates.
	In the following, let us demonstrate how this toolbox can be used for the quantum-state description in terms of phase-space distributions.

\section{Application: Unbalanced Homodyne Detection}
\label{Sec:UHD}

	In this section, we study the application of the developed technique to the problem of quantum-state reconstruction with the unbalanced homodyne detection (see Refs. \cite{Wallentowitz,Mancini}).
	Practical applications of this method require ideal photon-number determination, which is solely possible under special conditions \cite{Banaszek1999}.
	For instance, array detectors do not fall into this class of ideal measurement devices.
	In order to overcome this problem, it was proposed to reconstruct a click counterpart of the Cahill-Glauber $s$-parametrized phase-space quasi-probability distribution \cite{Luis, Lipfert2015}.
	Recently an experimental implementation of this approach was reported \cite{BohmannTielau}.
	Another approach utilizes the so-called fitting of data patterns \cite{Harder2014} for the local reconstruction of the Wigner function \cite{Harder2016}.
	Here, we show how the geometrical method can be applied to the reconstruction of $s$-parametrized phase-space quasiprobability distribution \cite{Cahill1,Cahill2} $P\left(\alpha;s\right)$ using array detectors and our error estimation.

\begin{figure}[htb]
	\includegraphics[width=1\linewidth]{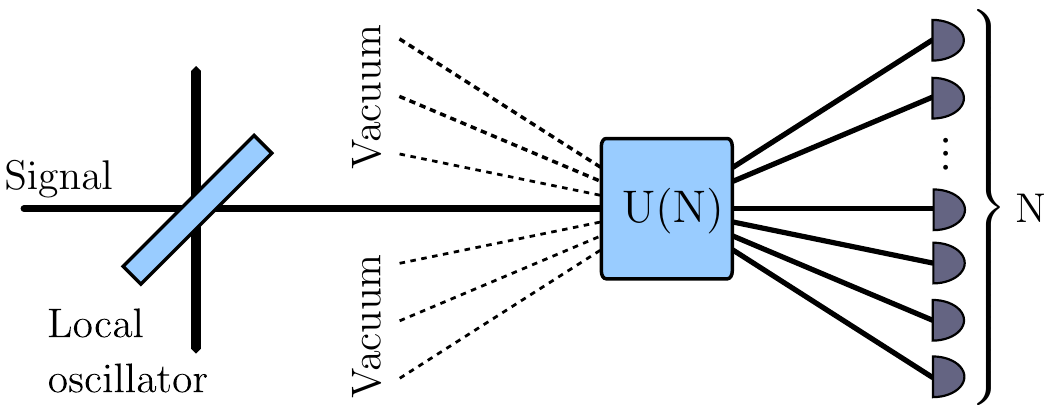}
	\caption{\label{Fig:UHD}
		Scheme for unbalanced homodyne detection \cite{Wallentowitz,Mancini}.
		The signal is combined with a local oscillator on a beam splitter with a transmission coefficient close to one.
		The output is then sent to an array detector (see also Refs. \cite{Luis,Lipfert2015}).
	}
\end{figure}

	The idea of the reconstruction consists in the fact that the Cahill-Glauber distribution $P\left(\alpha;s\right)$ can be presented as the expectation value of the operator
	\begin{align}
		\hat{P}\left(\alpha;s\right)=\frac{2}{\pi(1-s)}:\exp\left[-\frac{2}{1-s}\hat{n}(\alpha)\right]:,
		\label{Eq:UHD_Rec}
	\end{align}
	where
	\begin{align}
		\hat{n}(\alpha)=(\hat{a}^{\dag}-\alpha^\ast)(\hat{a}-\alpha)
	\end{align}
	is the displaced photon-number operator \cite{Wallentowitz}; $\hat{a}$ and $\hat{a}^\dag$ are field annihilation and creation operators, respectively.
	To get the value of the function $P\left(\alpha;s\right)$ at the point $\alpha$, one performs a displacement $-\alpha$ of the quantum state in phase space and then one measures the expectation value of the operator \eqref{Eq:UHD_Rec}, which is proportional to the normal-ordered exponent of the photon-number operator \eqref{Eq:ExpOpN}.
	With a minimal amount of loss, the displacement may be achieved by combining the signal field with the field of the local oscillator via a beam splitter with large transmission coefficient (see Fig. \ref{Fig:UHD}).
	For $s=-1,0,1$, the quasiprobability distribution $P\left(\alpha;s\right)=\langle\hat P\left(\alpha;s\right)\rangle$ is the Husimi-Kano function $Q\left(\alpha\right)$ \cite{Husimi, Kano}, the Wigner function $W\left(\alpha\right)$ \cite{Wigner}, and the Glauber-Sudarshan function $P\left(\alpha\right)$ \cite{Glauber,Sudarshan}, respectively.

	The expectation value of the operator $\hat{P}\left(\alpha;s\right)$ can be estimated from Eq. \eqref{Eq:ExpectEstim} for $\hat{B}=\hat{P}\left(\alpha;s\right)$.
	The corresponding contravariant coordinates are given by Eqs. \eqref{Eq:RisingIndeces}, \eqref{Eq:CovCoordExpr}, and \eqref{Eq:F}, for which
	\begin{align}
		F_{N;k}=\frac{2N}{\pi\left[N(1-s)+k(1+s)\right]}
		\label{Eq:F_for_UHD}
	\end{align}
	is easily obtained from Eq. \eqref{Eq:F_for_ExpN} by substituting $t=2/(1-s)$ and multiplying by $2/\pi(1-s)$.
	The HS norm of this operator is similarly obtained from Eq. \eqref{Eq:HSN_ExpN},
	\begin{align}
		\|\hat{P}\left(\alpha;s\right)\|_\mathrm{HS}^{[2]}=-\frac{1}{\pi^{[2]}s}
		\label{Eq:HSN_UHD}
	\end{align}
	for $s<0$.
	This norm can be used for calculating the corresponding HS mismatch.
	For $s\geq0$, the operator $\hat{P}\left(\alpha;s\right)$ does not belong to the HS class.
	However, a proper truncation of this operator according to Eq. \eqref{Eq:OpBExpMod} resolves this problem.
	In this case, the truncation parameter $M$ is chosen in such a way that the probability of more than $M$ photons at the detected output of the beam splitter is negligible.

\begin{figure}[htb]
	\includegraphics[width=\linewidth]{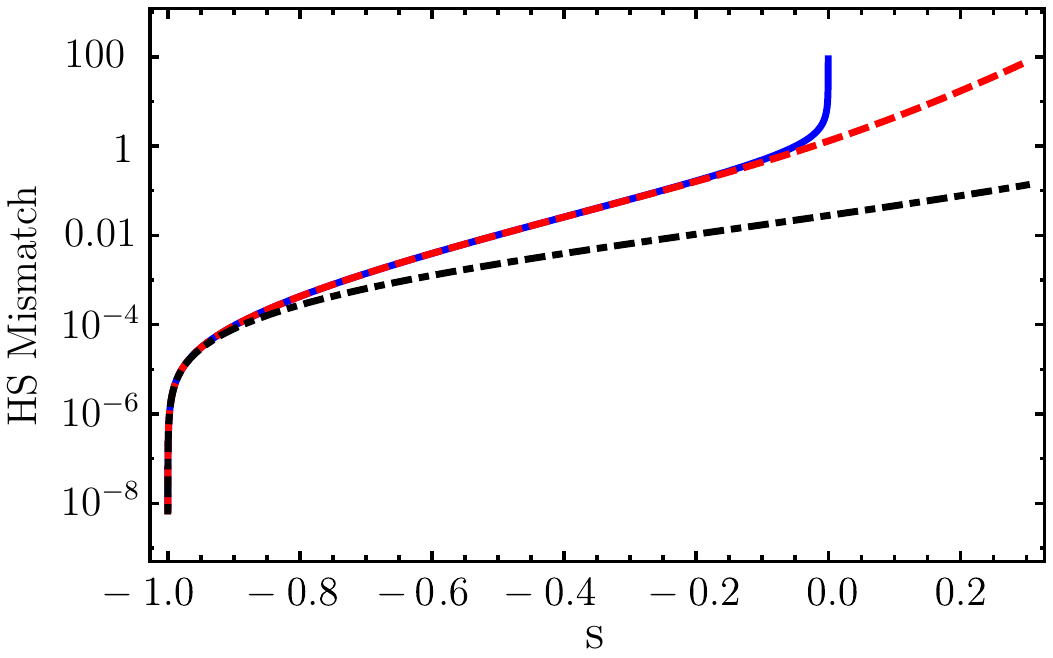}
	\caption{\label{Fig:Mismatch_UHD}
		The HS mismatch for the operator $\hat{P}\left(\alpha;s\right)$ (solid line) and for its truncations up to $M=7$ (dashed line) and $M=2$ (dot-dashed line) as functions of the parameter $s$.
	}
\end{figure}

	The HS mismatch for the operator $\hat{P}\left(\alpha;s\right)$ and its truncated versions are presented in Fig. \ref{Fig:Mismatch_UHD}.
	For small values of the parameter $s$, the maximal error is small too.
	However, when the value of $s$ is close to zero, the HS mismatch of the full operator increases rapidly.
	At the same time, the HS mismatches of truncated operators still have acceptable values for $s=0$ and even for larger $s$.
	This means that the reconstruction of states with a small photon number is also possible for such values of $s$.
	Note that with increasing $\alpha$, the effective truncation parameter $M$ increases too.
	This implies that the reconstructed Cahill-Glauber distributions with $s\gtrsim0$ gives larger noise for larger values of $|\alpha|$.

	In order to demonstrate the applicability of the method, we performed some numerical simulations.
	Specifically, we simulated the displaced click statistics $\varrho_n(\alpha)$ for the squeezed-vacuum state, whose Wigner function is given by
	\begin{align}
		W\left(\alpha\right)=\frac{2}{\pi}\exp\left[-\boldsymbol{\lambda}^\dag J V^{-1}J
		\boldsymbol{\lambda}\right],
		\label{Eq:WignerFSqVac}
	\end{align}
	where $\boldsymbol{\lambda}=\left(\begin{smallmatrix}\alpha\\\alpha^\ast\end{smallmatrix}\right)$, $J=\left(\begin{smallmatrix}1&0\\0&-1\end{smallmatrix}\right)$ is the symplectic matrix, and
	\begin{align}
		V=\left(\begin{array}{cc}
			\cosh 2\xi&\sinh 2\xi\\
			\sinh 2\xi&\cosh 2\xi
		\end{array}\right)
		\label{Eq:CovMatrSqVac}
	\end{align}
	is the covariance matrix, where $\xi$ is the squeezing parameter.
	For our example, we choose $\xi=0.8$, the number of detectors in the array is $N=8$, the efficiency is $\eta=0.7$, and we assume zero dark counts ($\nu=0$).
	We generated a sample size of $10^5$ data points for each value of $\alpha$.

\begin{figure}[htb]
	\includegraphics[width=1\linewidth]{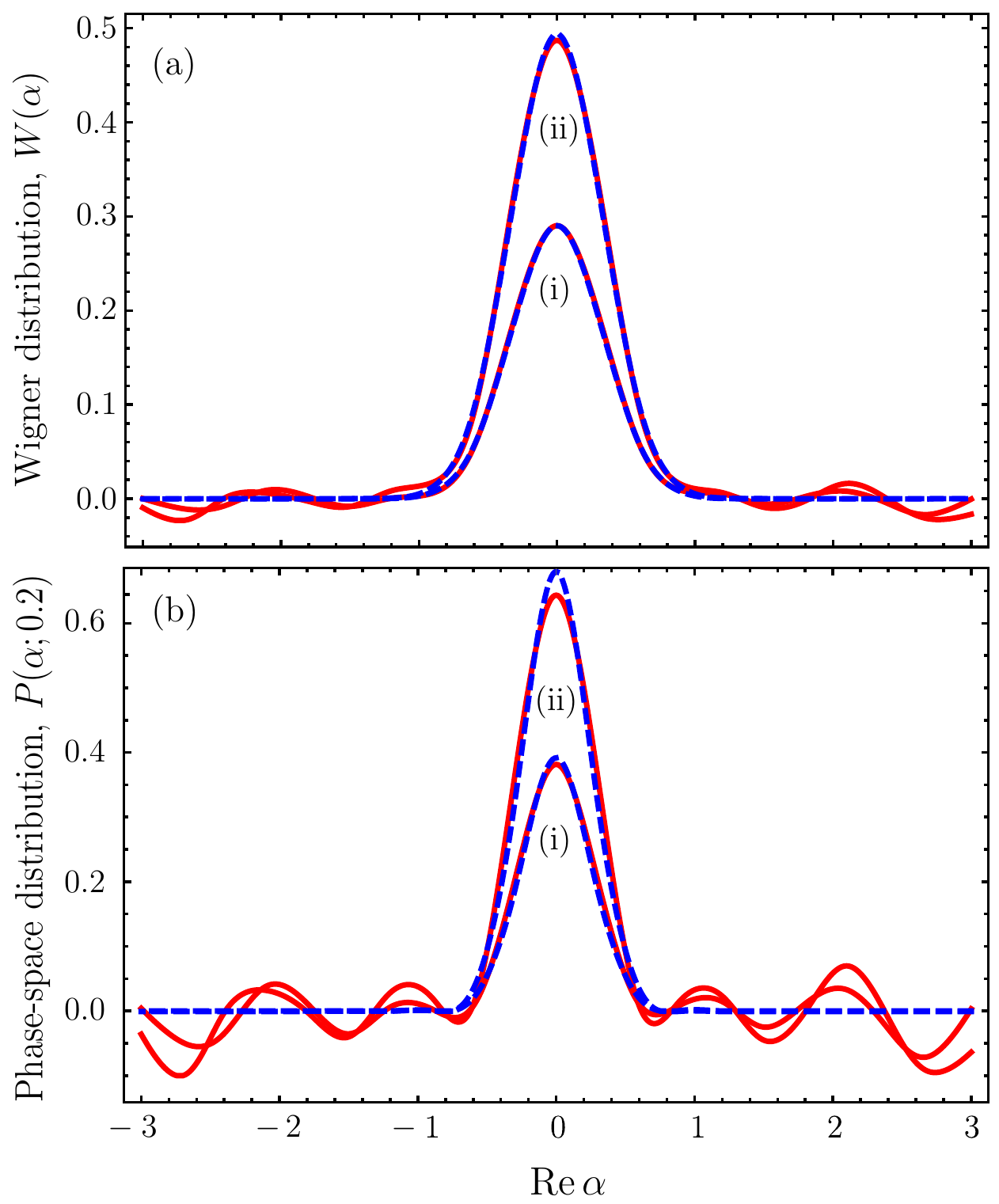}
	\caption{\label{Fig:Reconstruction_UHD}
		Theoretical (dashed lines) and reconstructed from the simulated data (solid lines) Cahill-Glauber phase-space distributions with (a) $s=0$ [Wigner function $W(\alpha)$] and (b) $s=0.2$ [distribution $P(\alpha;0.2)$] for a squeezed vacuum for (i) ${\rm Im}\,\alpha=1$ and (ii) ${\rm Im}\,\alpha=0$.
		Further details of the simulation are given in the text.
	}
\end{figure}

	The simulated data are substituted in Eq. \eqref{Eq:ExpectEstim} [see also Eqs. \eqref{Eq:RisingIndeces}, \eqref{Eq:CovCoordExpr}, and \eqref{Eq:F_for_UHD}] in order to reconstruct phase-space distributions with different values of $s$.
	The result is shown in Fig. \ref{Fig:Reconstruction_UHD} and compared with the ideal phase-space distributions.
	For a large range of values $s<0$, the reconstructed distribution fits the theoretical one with a negligible error.
	An acceptable result is specifically obtained for the Wigner function ($s=0$) [see Fig. \ref{Fig:Reconstruction_UHD}(a)].
	Moreover, for small values of the amplitude $\alpha$, we also obtain a good fit for positive $s$ [see Fig. \ref{Fig:Reconstruction_UHD}(b)].

\section{Summary and Conclusions}
\label{Sec:Conclusions}

	In summary, we devised a technique to study quantum-physical measurements which is based on a geometrical interpretation of Born's rule.
	In our framework, each (generalized) observable is a vector in the space of operators together with the corresponding POVM basis.
	We introduced the complementary concept of a contravariant operator-valued measure, the dual basis to the POVM under study.
	The contravariant coordinates of the observable define the possible measurement outcomes and covariant coordinates of the density operator define the probabilities of these outcomes.
	We have shown that other observables can be reconstructed or estimated from the statistics of the given measurement.
	In the latter case, the systematic error was determined and related to the orthogonal complement to the POVM.

	Our techniques are vital for the experimental determination of properties of quantum-optical systems measured with imperfect detectors.
	As an example, we considered the application to click-counting schemes, which consist of an array of individual on-off detectors.
	Based on our theory, we were able to reconstruct and quantitatively assess properties of the photon-number distribution without perfect photon-number resolution.
	For instance, the reconstruction of the photon-number statistics itself from the click statistics cannot be performed directly because of the large systematic error of the underlying ill-posed inversion problem.
	However, we successfully obtained expectation values for a number of other observables, which are functions of the photon-number operator.
	For many cases in such a scenario, we showed that our proposed geometrical technique robustly led to results with a small systematic error.

	As a second application, we considered state reconstruction problems.
	We demonstrated that our developed technique can be successfully used to obtain the Cahill-Glauber $s$-parametrized quasiprobability distribution by employing unbalanced homodyne detection.
	Typically, this experimental technique is challenging for implementations because of the need for a clear discrimination between adjacent photon numbers.
	Here, however, we have shown by our error estimation and numerical simulations that the unbalanced homodyne detection can be successfully implemented with an array of a few detectors only.
	It is also important to point out that, in general, this reconstruction does not require the application of regularization methods.
	Thus, our geometrical technique is an efficient tool for the quantum-state reconstruction.

	In conclusion, our concept of a geometrical interpretation of Born's rule provides useful tools for gaining a deeper understanding of the fundamental quantum measurement principle.
	In addition, we demonstrated that our approaches can be applied to many practical situations in which the determination of expectation values for different observables is needed.
	Specifically, our rigorous estimation of reconstruction errors is vital for the interpretation of experimental data.
	This includes the treatment of informationally incomplete detectors, which corresponds to a finite-dimensional POVM in a continuous-variable system.
	This was exemplified for photocounting via arrays of on-off detectors.
	Therefore, we presented an application-friendly theoretical approach to the geometry of quantum measurements in general and the photocounting in particular.

\begin{acknowledgments}
	The authors are grateful to M. Bohmann, D. Vasylyev, and S. Gerke for enlightening discussions and helpful comments.
	J.S. and W.V. acknowledge funding from the European Union's Horizon 2020 Research and Innovation program under Grant Agreement No. 665148 (QCUMbER).
\end{acknowledgments}

\appendix

\section{Application to photocounting}
\label{App:PD}

	Beyond click-counting schemes (see Sec. \ref{Sec:AD}), we present the application of our technique to the important example of photoelectric detection models here.
	In the ideal case, the standard form of Born's rule applies using the photon-number operator $\hat{n}$ with the orthonormal POVM $\hat{\Pi}_n=\ket{n}\bra{n}$.
	However, this is no longer true for realistic detectors, which is discussed in the following.

\subsection{Metric tensor and the COVM}

	The photocounting theory \cite{Mandel,Kelley} yields the imperfect detection of photons.
	The resulting POVM is defined in terms of the elements
	\begin{align}
		\hat{\Pi}_n=:\frac{\left(\eta\hat{n}+\nu\right)^{[n]}}{n!}\exp\left[-(\eta\hat{n}+\nu)\right]:.
		\label{POVM_Mandel}
	\end{align}
	As defined previously, $\eta$ is the detection efficiency, $\nu$ is the intensity of dark counts \cite{Semenov2008}, the superscript $[n]$ indicates the $n$th power of the operator $\eta\hat{n}+\nu$, and $:\cdots:$ denotes the normal-ordering prescription.
	We restrict ourselves to $\nu=0$.
	The generalization to arbitrary $\nu$ can be straightforwardly performed using the approaches in Refs. \cite{Semenov2008,Starkov}.
	Now we can expand the POVM elements in Eq. \eqref{POVM_Mandel} using photon-number eigenstates as
	\begin{align}
		\label{eq:LossyPhotons}
		\hat\Pi_n=&:\frac{(\eta\hat n)^{[n]}}{n!}\exp(-\eta\hat n):
		\\\nonumber
		=&\sum_{k=n}^\infty \binom{k}{n}\eta^{[n]}(1-\eta)^{[k-n]} |k\rangle\langle k|.
	\end{align}
	Furthermore, the measurement outcomes correspond to non-negative integers $C^n=n$.
	Hence, the resulting generalized observable is given by
	\begin{align}
		\hat{C}=\sum_n n\,\hat{\Pi}_n=\eta\hat{n}
		\label{Expanssion_Mandel}
	\end{align}
	For $\eta<1$, the eigenvalues of this operator $\eta n$ do not describe the actual number $n$.
	Therefore, we have a typical example of the generalized observable.

	To express the COVM, we compute the covariant metric tensor, which reads
	\begin{align}
		\label{MetricTensorMandel_Cov}
		&g_{nm}=\langle\hat\Pi_n,\hat\Pi_m\rangle_{\rm HS}
		\\
		=&\frac{\eta^{[n+m-1]}}{2-\eta}
		\sum_{k=\max[n,m]}^{n+m}
		\frac{k!}{(k-n)!(k-m)!(m+n-k)!}
		\nonumber
		\\
		&\phantom{\frac{\eta^{n+m-1}}{2-\eta}\sum_{k=\max\left[n,m\right]}^{n+m}}
		\times\frac{\left(1-\eta\right)^{[2k-n-m]}}{\eta^{[k]} \left(2-\eta\right)^{[k]}}.
		\nonumber
	\end{align}
	This metric tensor can be analytically inverted.
	This yields the contravariant metric tensor
	\begin{align}
		g^{nm}=\left(\frac{\eta-1}{\eta}\right)^{[n+m]}
		\sum_{k=0}^{\min[n,m]}
		\binom{n}{k}\binom{m}{k}\left(\eta-1\right)^{[-2k]}.
		\label{MetricTensorMandel_Contr}
	\end{align}
	Using Eq. \eqref{COVM}, we can also write the elements of the desired COVM in the form
	\begin{align}
		\hat{\Pi}^n=\left(\frac{\eta-1}{\eta}\right)^{[n]}
		:\exp(-\hat{n})\,{\rm L}_n \left(\frac{\hat{n}}{1-\eta}\right):,
		\label{COVM_Mandel}
	\end{align}
	where ${\rm L}_n$ denotes the $n$th Laguerre polynomial (see also Appendix \ref{App:MandelPC}).
	Let us stress that these operators are not positive semidefinite, which can be directly seen in the photon-number representation
	\begin{align}
		\label{eq:DualLossyPhoton}
		\hat\Pi^n=\sum_{k=0}^n \binom{n}{k}\left(\frac{1}{\eta}\right)^{[k]}\left(1-\frac{1}{\eta}\right)^{[n-k]}|k\rangle\langle k|,
	\end{align}
	where $1-1/\eta<0$ for $0<\eta<1$.

\subsection{Reconstructing statistical properties}

	Let us consider examples of contravariant coordinates for different (generalized) observables in the POVM basis \eqref{POVM_Mandel} to reconstruct the expectation value of these observables by measuring the outcomes of the photocounting with losses [see Eq. \eqref{Eq:Expansion}].

\begin{enumerate}[label=(\roman*)]
	\item
	Let us assume $\hat{B} = \hat{n}$, i.e., the photon-number operator.
	Then the corresponding coordinates read
	\begin{align}
		B^n=\langle\hat{\Pi}^n,\hat{n}\rangle_{\rm HS}=\frac{n}{\eta}.
	\end{align}
	This obvious relation not only indicates the proper function of our approach but also demonstrates that the mean number of photons is related to the mean number of photo-counts $\langle\hat{n}\rangle=\langle\hat{C}\rangle/\eta$.

	\item
	The contravariant coordinates of the operator $\hat{C}$ in Eq. \eqref{Expanssion_Mandel} are
	\begin{align}
		C^n=\langle\hat{\Pi}^n,\hat{C}\rangle_{\rm HS} =n.
	\end{align}
	Hence, although the eigenvalues $\eta n$ of the operator $\hat C$ do not describe the measurement outcomes, the contravariant coordinates do.

	\item
	To study the generating functions, let $\hat{B}=\exp(t\hat{n})$.
	Then we get
	\begin{align}
		B^n&=\langle\hat{\Pi}^n,\exp(t\hat{n})\rangle_{\rm HS}
		\nonumber\\
		&=\left(1+\frac{\exp(t)-1}{\eta}\right)^{[n]}.
		\label{Eq:Moments}
	\end{align}
	The Taylor expansion of this formula also leads to the next observation.

	\item
	Let us assume $\hat{B}=\hat{n}^{[m]}$.
	The mean value of this operator represents the corresponding $m$th moment.
	The contravariant coordinate of $\hat B$ reads 
	\begin{align}
		B^n=&\langle\hat{\Pi}^n,\hat{n}^{[m]}\rangle_{\rm HS}
		\nonumber\\
		=&\sum_{k=0}^{n}k^{[m]}\binom{n}{k}\frac{1}{\eta^{[k]}}\left(1-\frac{1}{\eta}\right)^{[n-k]}.
	\end{align}

	\item
	The normal-ordered generating function is given by the expectation value of $\hat{B}=:\exp(t\hat{n}):$.
	We find
	\begin{align}
		B^n=&\langle\hat{\Pi}^n,:\exp(t\hat{n}):\rangle_{\rm HS}
		=\left(1+\frac{t}{\eta}\right)^{[n]}.
	\end{align}
	Similar to Eq. \eqref{Eq:Moments}, we can expand it in Taylor series with respect to $t$ to obtain the next result.

	\item
	Let us assume $\hat{B}=:\hat{n}^{[m]}:$, yielding the normal-ordered $m$th moment.
	We have
	\begin{align}
		B^n=&\langle\hat{\Pi}^n,:\hat{n}^{[m]}:\rangle_{\rm HS}
		\nonumber\\
		=&\left\{\begin{array}{lcr}
			\frac{n!}{\eta^{[m]}\left(n-m\right)!}&\text{ for }&n\geq m\\
			0&\text{ for }&n<m.
			\end{array}\right.
	\end{align}

	\item
	Consider the contravariant coordinate of the POVM $\hat{\Pi}_m(0)=\ket{m}\bra{m}$ without loss.
	The transformation matrix to the lossy case is given by
	\begin{align}
		\label{Eq:MatrixTattenuation}
		&S_m{}^n=\langle\hat{\Pi}_m(0),\hat{\Pi}^n(\eta)\rangle_{\rm HS}\\
		=&\left\{\begin{array}{lcr}
			\binom{n}{m}\frac{1}{\eta^{[m]}}\left(1-\frac{1}{\eta}\right)^{[n-m]}&\text{ for }&n\geq m,\\
			0&\text{ for }&n<m.
		\end{array}\right.\nonumber
	\end{align}
	This straightforward consequence of our geometrical analysis resembles the results obtained in Refs. \cite{Lee2005,Herzog}.
	It allows for the transformation of imperfect photocounts into actual photon numbers.

	\item
	More generally, we can transform between arbitrary efficiencies $\eta\to\eta'$, which are described via the COVM elements $\hat\Pi^n(\eta)$ and POVM elements $\hat\Pi_n(\eta')$.
	The resulting transformation reads
	\begin{align}
		&[S(\eta^\prime,\eta)]_m{}^n=\langle\hat{\Pi}^n 
		\left(\eta\right),\hat{\Pi}_m \left(\eta^\prime\right)\rangle_{\rm HS}\label{Eq:eta_eta_prime}\\ 
		=&\left\{\begin{array}{lcr}
			\binom{n}{m}\frac{\eta^{\prime [m]}}{\eta^{[n]}}\left(\eta-\eta^\prime\right)^{[m-n]}&\text{ for }&n\geq m,\\
			0,&\text{ for }&n<m.
		\end{array}\right.\nonumber
	\end{align}
\end{enumerate}

	Therefore, we have the possibility to reconstruct expectation values and probability distributions of different observables with the technique developed from data obtained from imperfect photodetection.
	For instance, the above results can be used to infer nonclassical photon-number correlations in terms of normal-ordered moments \cite{Agarwal1992}.
	For certain situations, the methods considered cannot be directly applied, for example, when the experimental noise in the measured data affects the reconstruction.
	In such cases, one can complete the technique with regularization methods, e.g., the Landweber iteration \cite{Bertero, Aster}, which results in an acceptable accuracy of the reconstruction \cite{Starkov}.

\subsection{Metric tensors and the COVM for photocounting with losses}
\label{App:MandelPC}

	In the following, we describe methods of obtaining the co- and contrvariant metric tensors [see Eqs. \eqref{MetricTensorMandel_Cov} and \eqref{MetricTensorMandel_Contr}] as well as the COVM \eqref{COVM_Mandel} for the case of lossy photocounting.
	First of all, we present the equation for HS scalar product of two operators, which is useful for our calculations,
	\begin{align}
		{\rm Tr}(\hat{A}\hat{B})=
		\pi\int_\mathbb{C}{\rm d}^{[2]}\beta\, A^{C_Q}(\beta)B^{C_Q}(-\beta)\exp(|\beta|^{[2]}),
		\label{Eq:TraceFormula}
	\end{align}
	where we use the characteristic function of the $Q$ symbol of the operator
	\begin{align}
		A^{C_Q}(\beta)=\frac{1}{\pi^{[2]}}\int_\mathbb{C}{\rm d}^{[2]}\alpha\bra{\alpha}\hat{A}\ket{\alpha}\exp(\alpha\beta^\ast-\alpha^\ast\beta),
	\end{align}
	$B^{C_Q}(\beta)$ is defined similarly, and $\ket{\alpha}$ is the coherent state.
	For the POVM \eqref{POVM_Mandel} and $\nu=0$, we have $\bra{\alpha}\hat{\Pi}_n\ket{\alpha}=\exp(-\eta|\alpha|^{[2]})\eta^{[n]}\left|\alpha\right|^{[2n]}/n!$, which yields, for $g_{nm}={\rm Tr}[\hat{\Pi}_n\hat{\Pi}_m]$ the expression  \eqref{MetricTensorMandel_Cov}.

	Alternatively, the covariant metric tensor can be written as
	\begin{align}
		g_{nm}=\sum_{k=0}^{\infty}T_{n}{}^{k}T_{m}{}^{k}.
		\label{Eq:CovMetrTensViaT_losses}
	\end{align}
	Here $T_{n}{}^{k}$ can be obtained via the expansion coefficients in the series
	\begin{align}
		\hat{\Pi}_n=\sum_{k=0}^{\infty}T_{n}{}^{k}\ket{k}\bra{k}.
	\end{align}
	Remembering that the contravariant metric tensor $g^{nm}$ is the inverse of the covariant metric tensor $g_{nm}$, one gets from Eq. \eqref{Eq:CovMetrTensViaT_losses} that
	\begin{align}
		g^{nm}=\sum_{k=0}^{\infty}S_{k}{}^{n}S_{k}{}^{m},
		\label{Eq:ContrMetrTensViaT_losses}
	\end{align}
	where $S_{k}{}^{n}$ is given by Eq. \eqref{Eq:MatrixTattenuation} (see. Ref. \cite{Herzog,Lee2005}).
	After straightforward algebra, one finds the expression for the contravariant metric tensor in Eq. \eqref{MetricTensorMandel_Contr}. 

	The explicit form of the COVM is obtained by substituting Eq. \eqref{MetricTensorMandel_Contr} into Eq. \eqref{COVM}.
	This immediately results in the expression \eqref{eq:DualLossyPhoton} for the COVM.	
	In order to rewrite the COVM in the normal-ordered form, one has to calculate $\bra{\alpha}\hat{\Pi}^n\ket{\alpha}$ from Eq. \eqref{eq:DualLossyPhoton} and then replace $\alpha\mapsto\hat{a}$ and $\alpha^\ast\mapsto\hat{a}^\dag$ employing normal ordering.
	Eventually, this yields Eq. \eqref{COVM_Mandel}.

\section{Functions of generalized observables}
\label{App:Functions}

	If we consider the standard observable $\hat{A}$ with the measurement outcomes $A^n$, then the operator $F(\hat{A})$ represents the observable with the measurement outcomes $F(A^n)$.
	However, for the generalized observable $\hat{C}$, this is not true in general.
	For example, one could consider the click-number operator for the array detectors \cite{Sperling2013}.
	The square of this operator will not correspond to the observable for which we ascribe the value $k^{[2]}$ for the outcome with $k$ clicks.
	However, such an observable appears to be important for the verification of nonclassical light \cite{Sperling2012b}.

	Let us consider this question in more detail by starting with the special case for which $F(C^n)=(C^n)^{[2]}$.
	In close analogy with the star product in the phase-space representation of quantum mechanics \cite{Groenwold,Moyal,Zachos}, we say that the corresponding operator $\hat{C}^{[\star 2]}$ is star squared with respect to the POVM $\hat{\Pi}_n$,
	\begin{align}
		\hat{C}^{[\star 2]}=\sum_n (C^n)^{[2]}\hat{\Pi}_n=\sum_n 
		\langle\hat{\Pi}^n,\hat{C}\rangle_{\rm HS}^{[2]}\hat{\Pi}_n.
		\label{Eq:StarSquare}
	\end{align}
	It is worth mentioning that in the case of a standard observable, for which $\hat{\Pi}_n={\hat{\Pi}^n}=\ket{A_n}\bra{A_n}$, we immediately retrieve the standard relation $\hat{A}^{[\star 2]}=\hat{A}^{[2]}$.

	Similar relations can be formulated for any higher-order moment and, more generally, for any function of the observable $\hat{C}$.
	Thus, in a close analogy we present the star function of the observable with respect to the given POVM,
	\begin{align}
		F^{\star}(\hat{C})=\sum_n F(C^n)\hat{\Pi}_n=\sum_n 
		F(\langle\hat{\Pi}^n,\hat{C}\rangle_{\rm HS})\hat{\Pi}_n.
		\label{Eq:StarF}
	\end{align}
	Therefore, in the case of generalized observable, the star function has to be employed.

\section{Covariant metric tensor and covariant coordinates for array detectors}
\label{App:ArrayCov}

	Let us further elaborate the technique of calculation of the covariant metric tensor $g_{nm}$ for the case of array detectors and give additional expressions for the covariant coordinate, which includes detection losses with the efficiency $\eta$ and dark counts with the mean intensity $\nu$.
	These values can be obtained experimentally with the detector-calibration technique in Ref. \cite{Bohmann}.

	We start with the introduction of a more general expression than a metric tensor,
	\begin{align}
		\label{Eq:TrPOVM_Array}
		&\mathrm{Tr}\left[\hat{\Pi}_n\left(N,\eta,\nu\right)\hat{\Pi}_m\left(N^\prime,\eta^\prime,\nu^\prime\right)\right]\\
		=&\binom{N}{n}\binom{N^\prime}{m}\sum_{k=0}^{n}\sum_{l=0}^{m}\binom{n}{k}\binom{m}{l}(-1)^{[n+m-k-l]}F_{N,N^\prime;k,l},
		\nonumber
	\end{align}
	where
	\begin{align}
		&F_{N,N^\prime;k,l}\\
		=&\frac{NN^\prime \exp\left(-\nu\frac{N-k}{N}\right)\exp\left(-\nu\frac{N^\prime-l}{N^\prime}\right)}{N(N^\prime-l)\eta^\prime+N^\prime(N-k)\eta+\eta\eta^\prime(N-k)(N^\prime-l)}.\nonumber
	\end{align}
	Equation \eqref{Eq:TrPOVM_Array} represents the covariant coordinate of the POVM $\hat{\Pi}_m\left(N^\prime,\eta^\prime,\nu^\prime\right)$ in the basis of the POVM $\hat{\Pi}_n\left(N,\eta,\nu\right)$.
	This expression has been obtained by using the rule \eqref{Eq:TraceFormula}.
	The covariant metric tensor is obtained by setting $N=N^\prime$, $\nu=\nu^\prime$ and $\eta=\eta^\prime$.
	Particularly for $\nu=0$ and $\eta=1$, we obtain Eq. \eqref{Eq:CovMT}.

	The covariant coordinates for different observables are given by Eq. \eqref{Eq:CovCoordExpr}.
	However, the coefficients $F_{N;k}$ in the most general case are given by 
	\begin{align}
		F_{N;k}=\mathrm{Tr}\left[\hat{B}:\exp\left(-\frac{N-k}{N}g(\hat{n})\right):\right].
		\label{Eq:F_eta_nu}
	\end{align}
	In the following, we present the corresponding results for the families of operators studied in Sec. \ref{Sec:AD}.

\begin{enumerate}[label=(\alph*)]
	\item
	Let us assume $\hat{B}=\ket{m}\bra{m}$, which is a projector on a Fock state.
	In this case, the covariant coordinates can be found by expanding the POVM \eqref{Eq:POVMarray} with respect to $\ket{m}\bra{m}$,
	\begin{align}
		\label{TMatrix_eta_nu}
		B_n=&\Big[\ket{m}\bra{m}\Big]_n
		\\=&\binom{N}{n}\sum_{l=0}^{n}
		\binom{n}{l}(-1)^{[n-l]}\left(\frac{\eta l+N(1-\eta)}{N}\right)^{[m]}\nonumber\\ &\times\exp\left(-\nu\frac{N-l}{N}\right).\nonumber
	\end{align}

	\item
	Let us assume $\hat{B}=\exp(-t\hat{n})$.
	In this case $F_{N;k}$ in Eq. \eqref{Eq:CovCoordExpr} is given by
	\begin{align}
		F_{N;k}=\frac{N\exp\left(-\nu\frac{N-k}{N}\right)}{N[1-\exp(-t)]+\eta(N-k)\exp(-t)}.
		\label{Eq:F_for_Exp_eta_nu}
	\end{align}

	\item
	Let us assume $\hat{B}=\hat{n}^{[m]}$.
	The corresponding coefficients $F_{N;k}$ can be obtained via expanding Eq. \eqref{Eq:F_for_Exp_eta_nu} in a series with respect to $-t$.
	After some algebra, this yields
	\begin{align}
		F_{N;k}=&\exp\left(-\nu\frac{N-k}{N}\right)
		\label{Eq:Mom_eta_nu}\\
		&\times\left(x\frac{{\rm d}}{{\rm d} x}\right)^{[m]}(1-x)^{[-1]}\Big\rvert_{x=1-\eta (N-k)/N}.\nonumber
	\end{align}

	\item
	Let us assume $\hat{B}=:\exp(-t\hat{n}):$.
	The coefficients $F_{N;k}$ read
	\begin{align}
		F_{N;k}=\frac{N\exp\left(-\nu\frac{N-k}{N}\right)}{\eta(N-k)(1-t)+tN}.
		\label{Eq:F_for_ExpN_eta_nu}
	\end{align}

	\item
	Let us assume $\hat{B}=:\hat{n}^{[m]}:$.
	The coefficients $F_{N;k}$ are given by
	\begin{align}
		F_{N;k}=&\exp\left(-\nu\frac{N-k}{N}\right)
		\label{Eq:MomN_eta_nu}\\
		&\times\frac{m!N}{\eta(N-k)}\left(\frac{N}{\eta N-k}-1\right)^{[m]}.\nonumber
	\end{align}
\end{enumerate}

	By setting $\eta=1$ and $\nu=0$ in Eqs. \eqref{TMatrix_eta_nu}, \eqref{Eq:F_for_Exp_eta_nu}, \eqref{Eq:Mom_eta_nu}, \eqref{Eq:F_for_ExpN_eta_nu}, and \eqref{Eq:MomN_eta_nu}, one obtains Eqs. \eqref{TMatrix}, \eqref{Eq:F_for_Exp}, \eqref{Eq:Mom}, \eqref{Eq:F_for_ExpN}, and \eqref{Eq:MomN}, respectively.

\section{The pseudoinversion problem}
\label{App:Pseudoinv}

	Here we provide additional details on our geometrical technique in relation to the pseudoinversion problem.
	For this reason, we consider a finite-dimensional set of POVM operators $\hat{\Pi}_n$ for $\mathcal{I}=\{0,\ldots,N-1\}$ like that described for the array detector [see Eq. \eqref{Eq:POVMarray}].
	In addition, the infinite-dimensional set, i.e. $\mathcal{I}=\mathbb{N}_0$, of photocounting POVM [see Eq. \eqref{POVM_Mandel}] is denoted in the following by $\hat{\Lambda}_n$. 
	Considering $\hat{\Pi}_n$ as a given observable in the basis of $\hat{\Lambda}_n$, we can write, similar to the rule \eqref{Eq:Expansion},
	\begin{align}
	\hat{\Pi}_m=\sum\limits_{n=0}^{\infty}T_m{}^n\hat{\Lambda}_n.\label{Eq:ArrayPOVMviaPC_POVM}
	\end{align}
	Here,
	\begin{align}
		T_m{}^n=\langle\hat{\Lambda}^n,\hat{\Pi}_m\rangle_{\rm HS}
		\label{Eq:TransfMatr}
	\end{align}
	can be considered as the contravariant coordinate of the operator $\hat{\Pi}_m$ in the basis of operators $\hat{\Lambda}_n$ or the transformation matrix between different bases.
	The inverse relation
	\begin{align}
		\hat{\Lambda}_k=\sum\limits_{n=0}^{N-1}S_m{}^n\hat{\Pi}_n+\hat{R}_m
		\label{Eq:RelTransfMatrInv}
	\end{align}
	contains the orthogonal complement $\hat{R}_m$ because the basis of $\hat{\Pi}_k$ is finite.
	The transformation matrix is described by the components
	\begin{align}
		S_k{}^n=\langle\hat{\Pi}^n,\hat{\Lambda}_m\rangle_{\rm HS},
		\label{Eq:TransfMatrInv}
	\end{align}
	which can also be considered as the contravariant coordinate of the operator $\hat{\Lambda}_m$ in the basis of operators $\hat{\Pi}_n$.

	It is important to note the following features:
	\begin{align}
		T_m{}^n=\langle\hat{\Lambda}^n(1,0),\hat{\Pi}_m(1,0)\rangle_{\rm HS}=\langle\hat{\Lambda}^n(\eta,\nu),\hat{\Pi}_m(\eta,\nu)\rangle_{\rm HS}, \label{Eq:T_Inv}\\
		S_m{}^n=\langle\hat{\Pi}^n(1,0),\hat{\Lambda}_m(1,0)\rangle_{\rm HS}=\langle\hat{\Pi}^n(\eta,\nu),\hat{\Lambda}_m(\eta,\nu)\rangle_{\rm HS}. \label{Eq:S_Inv}
	\end{align}
	In those relations, we explicitly introduce the dependence on the detection efficiency $\eta$ and dark count intensity $\nu$ in POVM and COVM operators.
	This property means that the actual matrices $T_m{}^n$ and $S_m{}^n$ do not depend on $\eta$ and $\nu$.
	Due to the orthogonality of the POVM elements $\hat{\Lambda}_n(1,0)$ [i.e. $\langle\hat{\Lambda}_n(1,0),\hat{\Lambda}_m(1,0)\rangle_{\rm HS}=\delta_{nm}$], the index $n$ in the matrix $T_m{}^n$ can be lowered $T_m{}^n=T_{mn}$.
	This yields that the explicit form for $T_m{}^n$ coincides with the covariant coordinate $B_n$, of the operator $\ket{m}\bra{m}$ given by Eq. \eqref{Eq:CovCoordExpr}. 

	In order to simplify notation, we introduce the symbols for the matrices $\boldsymbol{T}=(T_m{}^n)_{m\in\{0,\ldots, N-1\}, n\in\mathbb{N}_0}$ and $\boldsymbol{S}=(S_m{}^n)_{m\in\mathbb{N}_0, n\in\{0,\ldots, N-1\}}$.
	By using Eqs. \eqref{Eq:TransfMatr}, \eqref{Eq:TransfMatrInv}, \eqref{Eq:T_Inv}, and \eqref{Eq:S_Inv} as well as the rule of rising indices \eqref{Eq:RisingIndeces}, one can connect the matrices $\boldsymbol{T}$ and $\boldsymbol{S}$,
	\begin{align}
		\boldsymbol{S}=\boldsymbol{T}^\mathrm{T}\left(\boldsymbol{T}\boldsymbol{T}^\mathrm{T}\right)^{[-1]}.
	\end{align} 
	This means that $\boldsymbol{S}$ is the Penrose-Moore pseudoinverse \cite{Ben-Israel} of $\boldsymbol{T}$, which is often written as $\boldsymbol{S}=\boldsymbol{T}^+$.

	Let us mention some important properties of the Penrose-Moore pseudoinverse in connection to our application.
	First, we introduce vectors of the click statistics $\boldsymbol{\varrho}=(\varrho_m)_{m\in\{0,\ldots, N-1\}}$ and the photon-number statistics $\boldsymbol{p}=(p_n)_{n\in\mathbb{N}_0}$; recall that $\varrho_m=\langle\hat{\varrho},\hat{\Pi}_m \rangle_{\rm HS}$ and $p_n=\langle\hat{\varrho},\hat{\Lambda}_n \rangle_{\rm HS}$.
	Applying the quantum-state functional to Eq. \eqref{Eq:ArrayPOVMviaPC_POVM}, one gets the transformation relation
	\begin{align}
		\boldsymbol{\varrho}=\boldsymbol{T}\boldsymbol{p}.
	\end{align}
	Similarly, introducing an approximate photon-number statistics via the vector $\boldsymbol{\tilde{p}}=(\tilde p_n)_{n\in\mathbb{N}_0}$, where $\tilde{p}_n=\langle\hat{\varrho},\hat{\Lambda}_n-\hat{R}_n \rangle_{\rm HS}$, Eq. \eqref{Eq:RelTransfMatrInv} is rewritten as
	\begin{align}\label{Eq:ApprPNStat}
		\boldsymbol{\tilde{p}}=\boldsymbol{S}\boldsymbol{\varrho}.
	\end{align}
	This means that the approximate vector \eqref{Eq:ApprPNStat} minimizes the functional
	\begin{align}\label{Eq:LSqFunktional}
		\mathcal{F}=\| \boldsymbol{T}\boldsymbol{\tilde{p}}-\boldsymbol{\varrho}\|^{[2]},
	\end{align}
	where $\|\cdots\|$ is the $L2$-norm of the vector \cite{Ben-Israel}.
	Therefore, our geometrical method gives the best approximate solution in the sense of least squares, which is known to be described via the Penrose-Moore pseudoinverse.
	Still, the direct applications of the pseudoinversion to the problem of photon-number reconstruction can lead to unacceptable results (see Sec. \ref{Sec:PNStat}).
	For this reason, regularization methods can be introduced by applying a modification of the functional \eqref{Eq:LSqFunktional} \cite{Bertero,Aster}.

	As mentioned previously, the matrix $\boldsymbol{S}$ is the so-called Penrose-Moore pseudoinverse to the matrix $\boldsymbol{T}$, which has to satisfy the conditions:
	(i) $\boldsymbol{S}\boldsymbol{T}\boldsymbol{S}=\boldsymbol{T}$,
	(ii) $\boldsymbol{T}\boldsymbol{S}\boldsymbol{T}=\boldsymbol{S}$,
	(iii) $(\boldsymbol{T}\boldsymbol{S})^\mathrm{T}=\boldsymbol{T}\boldsymbol{S}$, and
	(iv) $(\boldsymbol{S}\boldsymbol{T})^\mathrm{T}=\boldsymbol{S}\boldsymbol{T}$.
	Specifically, the conditions (iii) and (iv) reflect the connection to the HS space.
	It is also worth mentioning that beyond the HS structure, other types of pseudoinverses can be introduced.
	An example is the matrix $\boldsymbol{\tilde{S}}$, given by
	\begin{align}
		\label{Eq:IncomplPsInv}
		\tilde{S}_m{}^n=\binom{N}{n}^{[-1]}\frac{N^{[m]}}{n!}\left(-1\right)^{[m-n]}
		\left[\begin{array}{c}
		n\\m
		\end{array}\right],
	\end{align}
	where $\left[\begin{smallmatrix}n\\m\end{smallmatrix}\right]$ are Stirling numbers of the first kind (see also the Supplemental Material to Ref. \cite{Kroeger}).
	This type of pseudoinverse does not satisfy condition (iv).


\end{document}